\documentclass[fleqn,usenatbib]{mnras}

\usepackage[T1]{fontenc}

\DeclareRobustCommand{\VAN}[3]{#2}
\let\VANthebibliography\thebibliography
\def\thebibliography{\DeclareRobustCommand{\VAN}[3]{##3}\VANthebibliography}

\usepackage{graphicx}	
\usepackage{amsmath}	
\usepackage{amssymb}	
\usepackage{siunitx}

\usepackage{filecontents}
\usepackage{subfigure}
\usepackage{tikz}
\usepackage{multirow}
\usepackage{xcolor}
\usepackage{multicol}
\usepackage{hyperref}
\usepackage{gensymb}

\newcommand{\Msun}{\mbox{M$_{\odot}$}}

\newcommand{\Teff}{\mbox{$T_\mathrm{eff}$}}

\newcommand{\EW}{\mbox{$W_{\textnormal{$\lambda$}}$}}
\newcommand{\angstrom}{\text{\normalfont\AA}}

\title[Weak magnetism detected in WD\,J1653$-$1001]{Detection of a weak magnetic field in the Balmer emission line white dwarf WD\,J1653$-$1001}

\author[A. K. Elms]{Abbigail K. Elms$^{1}$\thanks{Contact e-mail: \href{mailto:Abbigail.K.Elms@warwick.ac.uk}{Abbigail.K.Elms@warwick.ac.uk}},
{Stefano Bagnulo$^{2}$,}
{Pier-Emmanuel Tremblay$^{1}$,}
{Tim Cunningham$^{3}$,}
{James Munday$^{1}$,}
\newauthor
{John Landstreet$^{4}$,}
{Kareem El-Badry$^{5}$,}
{Ilaria Caiazzo$^{5,6}$,}
{Carl Melis$^{7}$,}
{Viktoria Pinter$^{8,9}$, and}
\newauthor
{Alycia Weinberger$^{10}$}
\\
$^{1}$Department of Physics, University of Warwick, Coventry, CV4 7AL, UK\\
$^{2}$Armagh Observatory \& Planetarium, College Hill, Armagh BT61 9DG, UK\\
$^{3}$Center for Astrophysics | Harvard \& Smithsonian, 60 Garden St., Cambridge, MA 02138, USA\\
$^{4}$Department of Physics \& Astronomy, University of Western Ontario, London, Ontario N6A 3K7, Canada\\
$^{5}$Division of Physics, Mathematics and Astronomy, California Institute of Technology, Pasadena, CA91125, USA\\
$^{6}$Institute of Science and Technology Austria, Am Campus 1, 3400, Klosterneuburg, Austria\\
$^{7}$Department of Astronomy \& Astrophysics, University of California San Diego, La Jolla, CA 92093-0424, USA\\
$^{8}$Centro Astronómico Hispano en Andalucía, Observatorio de Calar Alto, Sierra de los Filabres, 04550 Gérgal, Spain\\
$^{9}$Issac Newton Group of Telescopes, Calle Alvarez Abreu, 70, 38700 Santa Cruz de La Palma, Spain\\
$^{10}$Earth and Planets Laboratory, Carnegie Institution for Science, 5241 Broad Branch Rd NW, Washington, DC 20015, USA\\
}

\date{Accepted XXX. Received YYY; in original form ZZZ}

\pubyear{\the\year{}}

\usepackage{newtxtext,newtxmath}

\begin{document}
\label{firstpage}
\pagerange{\pageref{firstpage}--\pageref{lastpage}}
\maketitle

\begin{abstract}
The small DAHe and DAe spectral classes comprise isolated, hydrogen-dominated atmosphere white dwarfs that exhibit variable photometric flux and Balmer line emission. These mysterious systems offer unique insight into the complex interplay between magnetic fields, stellar rotation and atmospheric activity in single white dwarfs. DAHe stars have detectable magnetic fields through Zeeman-split spectral lines, whereas DAe stars lack such splitting. We report the first discovery and characterisation of magnetism in the DAe white dwarf WD\,J165335.21$-$100116.33 with new time-resolved spectropolarimetry from FORS2. We detect a weak but variable longitudinal magnetic field with values $\langle B_z \rangle > -9.2 \pm 2.4$\,kG and $\langle B_z \rangle < -2.2 \pm 1.0$\,kG. Independent ZTF and ATLAS photometry reveal a consistent period of $P = 80.3070 \pm 0.0007$\,h. Time-resolved optical spectroscopy obtained with six ground-based instruments demonstrates strong modulation in the strength of the H$\alpha$ and H$\beta$ Balmer line emission with $P = 80.2922 \pm 0.0108$\,h. The photometric flux and Balmer emission strength vary in antiphase, with the strongest magnetic detections coinciding with phases of low photometric flux and strong line emission. These characteristics support the theory that a magnetically active, temperature-inverted spot/region is producing an optically thin chromospheric emission region. Comparison with other DAe and DAHe white dwarfs reveals all systems have a strikingly similar antiphase phenomenology, reinforcing the theory that they are subject to a unified physical mechanism. With the detection of a weak magnetic field, we reclassify WD\,J165335.21$-$100116.33 as a low-field DAHe white dwarf. 
\end{abstract}

\begin{keywords}
white dwarfs -- stars: interiors -- stars: individual (WD\,J165335.21$-$100116.33) -- methods: observational -- methods: data analysis
\end{keywords}



\section{Introduction}
Over the past decade, the \textit{Gaia} mission \citep{Gaia2016, Gaia2022} has observed over two billion stars and other celestial objects in our Galaxy. From these observations, over $\approx 350\,000$ high-confidence white dwarf candidates have been identified \citep{GF2021}. Of these, 28 white dwarfs have been found to exhibit a rare combination of spectral characteristics and have been classified as DAHe or DAe white dwarfs \citep{Greenstein1985, Reding2020, Tremblay2020, Gansicke2020, Walters2021, OBrien2023, Reding2023, Manser2023}. These degenerate (`D') stars possess hydrogen-dominated atmospheres (`A'), Balmer line emission (`e'), and in many cases exhibit magnetic fields detectable via Zeeman-split emission line triplets (`H'). Two of these white dwarfs - WD\,J041246.84$+$754942.26 \citep[hereafter WD\,J0412$+$7549;][]{Tremblay2020} and WD\,J165335.21$-$100116.33 \citep[hereafter WD\,J1653$-$1001;][]{OBrien2023} - are spectroscopically classified as DAe in the literature, as they do not exhibit Zeeman-splitting but show an emission component at the centres of their Balmer lines. The remaining 26 stars are classified as DAHe. While DAHe stars exhibit surface magnetic field strengths ranging from $B\simeq 5 - 147$\,MG \citep{Greenstein1985, Gansicke2020, Reding2020, Walters2021, Manser2023, Reding2023}, upper limits for the two DAe white dwarfs were constrained to $B < 0.05$\,MG \citep{Elms2023}. 

The acquisition and analysis of time-resolved spectroscopic and photometric data of both classes, grouped together as DA(H)e white dwarfs, is an ongoing endeavor. Measured variability periods among these stars range from $\simeq 0.08 - 80$\,h, making complete phase spectroscopic coverage challenging for some stars. For instance, \citet{Elms2023} found WD\,J0412$+$7549 to have a period of $2.289114 \pm 0.000002$\,h and acquired full-phase spectroscopic coverage, whereas WD\,J1653$-$1001 had a tentative ZTF period of $80.534 \pm 0.087$\,h and only $2.4$ per cent phase coverage. Available survey data and targeted spectroscopic follow-up of DA(H)e white dwarfs have revealed that 57 per cent are photometrically variable and 29 per cent are spectroscopically variable \citep{Gansicke2020, Reding2020, Walters2021, Manser2023, Reding2023, Elms2023} - although these are likely lower limits due to insufficient instrumental sensitivity and time-domain coverage for some stars. For those DA(H)e white dwarfs that exhibit both types of variability, an antiphase relationship has been found between the photometric (flux) and spectroscopic (emission) maxima \citep{Gansicke2020, Reding2020, Walters2021, Manser2023, Reding2023, Elms2023}. 

DA(H)e white dwarfs are of particular astrophysical interest as an explanation for the physical origin of their characteristics remains elusive. They are apparently isolated stars and have a remarkable homogeneity in atmospheric parameters, with effective temperatures $7400\,\mathrm{K} \lesssim \Teff \lesssim 8500$\,K and white dwarf masses $0.5\,\Msun \lesssim \mathrm{M}_{\mathrm{WD}} \lesssim 0.8$\,\Msun. Also, they closely cluster in one region of the white dwarf cooling track at late cooling times on the \textit{Gaia} Hertzsprung–Russell diagram \citep[HRD;][]{Gansicke2020, Walters2021, Manser2023, Elms2023}, suggesting that DA(H)e stars may represent a short-lived evolutionary phase during which magnetic fields of vastly varying strength generate Balmer line emission. Therefore, it is likely that intrinsic mechanisms, such as structural evolution and/or internal dynamics, are the reason for their observed properties. The most direct scenario for the observations is that these stars have a photospheric dark spot/region with a temperature-inverted and optically thin chromospheric emission region \citep{Walters2021}. One possible explanation for this behaviour is that DA(H)e stars have a slowly emerging magnetic field \citep{Bagnulo2022,Camisassa2024,Moss2025}. This could be from a convective dynamo driven by white dwarf core crystallisation occurring at a specific time in the cooling sequence \citep{Ginzburg2022,Lanza2024}, although this interpretation remains uncertain \citep{Fuentes2023,Blatman2024,CastroTapia2024b} and some DA(H)e have not yet begun crystallisation \citep{Manser2023, Elms2023}. Additionally, these stars could have a remnant fossil field from earlier evolutionary phases that was previously buried beneath the surface and is now re-emerging through diffusion \citep{Braithwaite2004, Tout2004, Wickramasinghe2005, Camisassa2024, CastroTapia2025}. 

It is important to decipher the magnetic field strength of DA(H)e white dwarfs as magnetic fields are key drivers of structure and variability in stellar atmospheres. If DAe and DAHe white dwarfs share a common surface mechanism, it must explain line emission for both low ($< 0.05$\,MG) and high ($> 5$\,MG) magnetic fields. 

Since WD\,J0412$+$7549 and WD\,J1653$-$1001 show no evidence of Zeeman-split emission line triplets in their spectra, they either do not possess magnetic fields or their fields are too weak to be observationally detected via spectroscopy. \citet{Walters2021} mentioned that WD\,J0412$+$7549 should be magnetic at a detectable level. Spectropolarimetry offers a powerful method, capable of combining spectroscopy and polarimetry to measure the circular polarisation of light caused by a weak magnetic field of $\approx 1$\,kG~$- 1$\,MG \citep{Kawka2007} as a function of wavelength \citep[e.g.][]{LandstreetAngel1975, Friedrich1996, Vornanen2013}. This yields the mean longitudinal field ($\langle B_z \rangle$), which is the average magnetic field strength over the stellar hemisphere along the line-of-sight at the time of observation. Since magnetic field strength can differ from the measured $\langle B_z \rangle$ at other points on the white dwarf surface, time-series spectropolarimetric observations over the white dwarf spin period are necessary to constrain the field geometry and true surface field strength. A non-zero $\langle B_z \rangle$ constitutes a definitive detection of a magnetic field, enabling the measurement of field strengths that are too weak to produce detectable Zeeman splitting. 

In this work, we present new time-resolved spectroscopic and spectropolarimetric observations of the white dwarf WD\,J1653$-$1001 and analyse its time-series observations from the Zwicky Transient Facility \citep[ZTF;][]{Bellm2019, Masci2019} and the Asteroid Terrestrial-impact Last Alert System \citep[ATLAS;][]{Tonry2018}. Section~\ref{sec:observations} describes the observational data sets, while Section~\ref{sec:Analysis} details the analysis completed to investigate the white dwarf spin period, photometric and spectroscopic line variability, and $\langle B_z \rangle$. We discuss the implications of our results in Section~\ref{sec:Discussion} and present our conclusions in Section~\ref{sec:Conclusions}.

\section{Observations and data}
\label{sec:observations}


\begin{table*}
    \centering
	\caption{Astrometry from \textit{Gaia} DR3 and derived parameters from \citet{Elms2023}. Atmospheric parameters were calculated by performing weighted 3D spectroscopic fits. Quantities marked with an asterisk (*) are derived in this work: the polar magnetic field strength ($B_d$) is calculated from the 2024 FORS2 spectropolarimetric observations (Sections~\ref{sec:FORS2}~and~\ref{sec:Spectropolametric analysis}) and the spin period is the best-fitting period from the simultaneous $g$-, $r$-, $c$- and $o$-band MCMC (Section~\ref{sec:Photometric variability of WDJ1653}). Values are given in the J2016.0 epoch.}
	\label{tab:stellar parameters}
	\begin{tabular}{llc}
		\hline
		\hline
		Parameter &  & WD\,J1653$-$1001 \\
		\hline
		\hline
        Designation & & \textit{Gaia} DR3 4334641562477923712 \\
		RA & & 16:53:35.21\\
		Dec & & $-$10:01:16.33 \\
		Parallax & $\varpi$ [mas] & 30.65 $\pm$ 0.04 \\
		Distance & \textit{d} [pc] & 32.63 $\pm$ 0.04 \\ 
		Proper motion & $\mu_\alpha$ [mas\,yr$^{-1}$] & 159.38 $\pm$ 0.05 \\	
		 & $\mu_\delta$ [mas\,yr$^{-1}$] & $-$211.01 $\pm$ 0.03 \\
		Absolute magnitude & $M_{\rm G}$ [mag] & 13.140 $\pm$ 0.003 \\
		\hline
		Effective temperature & \Teff\ [K] & 7613 $\pm$ 95\\
		Surface gravity & $\log g$ [cm\,s$^{-2}$] & 7.893 $\pm$ 0.030 \\
		Mass & $\mathrm{M}_{\mathrm{WD}}$ [\Msun] & 0.53 $\pm$ 0.02 \\
		Radius & $R$ [$\times$10$^{-5}$ R$_\odot$] & 1366 $\pm$ 26 \\
		Cooling age & $\tau$ [Gyr] & 1.019 $\pm$ 0.050 \\
		Polar magnetic field strength* & $B_d$ [kG] & $22^{+32}_{-7}$ \\
		Spin period* & \textit{P} [h] & 80.3070 $\pm$ 0.0007  \\
		\hline
	\end{tabular}
\end{table*}

Table~\ref{tab:stellar parameters} presents astrometry for WD\,J1653$-$1001 from \textit{Gaia} Data Release 3 (DR3) and derived atmospheric parameters from 3D spectroscopic modeling performed in \citet{Elms2023}. The polar magnetic field strength and spin period are parameters calculated in this work, providing an important update from \citet{Elms2023}. 

\subsection{Time-domain spectroscopy and spectropolarimetry}
\label{sec:Time-domain spectroscopy and spectropolarimetry}

\begin{table*}
    \centering
	\caption{Time-domain spectroscopic observations for WD\,J1653$-$1001 obtained from ground-based telescopes, detailing the exposure time ($t_{\rm exp}$), number of exposures ($n_{\rm exp}$) for each observing run and the duration of the observing run. Numbers separated by a colon represent exposures taken in the blue:red arms. Observations are in ascending date order, with the UT date and time given at mid-exposure. Four exposures were taken with Binospec on 2025-05-20, however we only use two in this work as the latter two were low S/N.}
	\label{tab:observations}
	\begin{tabular}{cccccc}
		\hline
		\hline
		Date at mid-exposure & Time at mid-exposure & Telescope/Instrument & $t_{\rm exp}$ & $n_{\rm exp}$ & Duration \\
		yyyy-mm-dd & hh:mm:ss &  & [s] &  & [h] \\
		\hline
		\hline
		2018-05-22 & 08:43:45 & Shane/Kast & 3000:1000 & 1:3 & 0.83 \\
        2022-05-15 & 03:39:20 & VLT/FORS2 & 800 & 4 & 0.89 \\
        2022-06-01 & 02:30:11 & VLT/FORS2 & 800 & 4 & 0.89 \\
        2023-05-15 & 09:56:59 & Shane/Kast & 2000:1000 & 2:4 & 1.11 \\
        2023-05-28 & 04:16:21 & Magellan/MIKE & 1200 & 1 & 0.33 \\
        2023-06-19 & 23:59:07 & INT/IDS & 1200 & 12 & 3.75 \\
        2023-06-25 & 05:57:27 : 05:52:26 & Shane/Kast & 2800:1100 & 1:3 & 0.78:0.93 \\
        2023-07-19 & 00:16:37 & INT/IDS & 1200 & 8 & 2.63 \\
        2023-09-21 & 00:02:00 & Magellan/MIKE & 1200 & 2 & 0.67 \\
        2024-05-18 & 03:25:09 & VLT/FORS2 & 330 & 8 & 0.73 \\
		2024-06-06 & 00:59:30 & VLT/FORS2 & 330 & 8 & 0.73 \\
		2024-06-16 & 00:59:54 & VLT/FORS2 & 330 & 8 & 0.73 \\
        2024-07-04 & 02:25:18 & VLT/FORS2 & 330 & 8 & 0.73 \\
        2024-07-11 & 01:56:24 & Magellan/MagE & 1200 & 3 & 1.00 \\
        2024-07-14 & 00:04:38 & VLT/FORS2 & 330 & 8 & 0.73 \\
		2024-07-15 & 00:05:22 & VLT/FORS2 & 330 & 8 & 0.73 \\
        2025-05-19 & 07:51:29 & MMT/Binospec & 1200 & 6 & 2.00 \\
        2025-05-20 & 10:37:41 & MMT/Binospec & 1200 & 2 & 0.67 \\
        2025-05-21 & 08:42:06 & MMT/Binospec & 1200 & 6 & 2.00 \\
        2025-05-22 & 09:44:29 & MMT/Binospec & 1200 & 5 & 1.67 \\
        \hline
	\end{tabular}
\end{table*}

Spectroscopic observations of WD\,J1653$-$1001 were made using five different ground-based telescopes and six instruments spanning seven years. The long time-frame between observations allows for a dedicated search for variability in the Balmer emission lines. Observational details are listed in Table~\ref{tab:observations} for WD\,J1653$-$1001, including the exposure times ($t_{\rm exp}$), number of exposures ($n_{\rm exp}$) and the total duration of each observing run. Spectropolarimetry for WD\,J1653$-$1001 was obtained with the FORS2 instrument spanning two years. Sections~\ref{sec:FORS2}~--~\ref{sec:Binospec} discuss the observations made with each telescope in detail. Two epochs of archival spectroscopic data (Section~\ref{sec:Kast}) previously presented in \citet{Elms2023} are re-analysed in this work, while all other observations are presented here for the first time. 

\subsubsection{VLT/FORS2 spectroscopy and spectropolarimetry}
\label{sec:FORS2}
Spectropolarimetric observations of WD\,J1653$-$1001 were obtained with the multi-mode optical instrument FOcal Reducer/low dispersion Spectrograph 2 \citep[FORS2; ][]{Appenzeller1998}, which is mounted on the Cassegrain focus of the UT1 telescope at the Very Large Telescope (VLT). The VLT is operated by the European Southern Observatory (ESO) and is located on the Cerro Paranal mountain, Chile. FORS2 provides wavelength-dependent total intensity spectra of a target, in addition to polarimetric measurements.

WD\,J1653$-$1001 was observed in two different observing runs with FORS2. The first observing run was in 2022, with four 800\,s exposures taken on 2022-05-15 and 2022-06-01. The standard resolution collimator (COLL\_SR) and 1200B grism (GRIS\_1200B+97) were used; this setup yields a 24.0\,\AA/mm dispersion and no order sorting filter was used. The wavelength range of the observations are $3500$~--~$5300$\,\AA\ with a central wavelength of $4350$\,\AA, therefore coverage of H$\beta$ was obtained. The default two $2000 \times 4000$\,pixel MIT CCDs were used with the $2 \times 2$ binned readout mode. As a slit width of $1.0$" was used, there was sufficient sampling with this binned readout mode.

The second observing run was in 2024, with eight 330\,s exposures taken on six nights from May to July. The COLL\_SR and 1200R grism (GRIS\_1200R+93) together with the GG435 order separation filter were used, yielding a 25.0\,\AA/mm dispersion. The observations have a wavelength range of $5750$~--~$7310$\,\AA\ and a central wavelength of $6500$\,\AA, therefore coverage of H$\alpha$ was obtained. The default two $2000 \times 4000$\,pixel MIT CCDs were used with the $2 \times 2$ binned readout mode. 

The polarimetric mode of FORS2 was employed for both 2022 and 2024 setups, providing circular spectropolarimetry across the optical range of WD\,J1653$-$1001. The reduced data yield the Stokes $I$ (total intensity), Stokes $V$ (circular polarisation), and the null profile \citep[$N_V$; ][]{Donati1997}, which is representative of the noise of the fraction of circular polarization ($V/I$). In essence, $N_V$ is a diagnostic tool to assess the reliability of the $V$ measurements and detect possible spurious signals. These data allow the detection of small magnetic fields through the measurement of $\langle B_z \rangle$ and its associated null field ($\langle N_z \rangle$) value, as described in Section~\ref{sec:Spectropolametric analysis}.

\subsubsection{Shane/Kast spectroscopy}
\label{sec:Kast}
The Kast Double Spectrograph on the Shane 3\,m Telescope at the Lick Observatory in California, USA, was used to observe WD\,J1653$-$1001 over three epochs. We utilised the default set-up of the Kast spectrograph, using a Fairchild $2000 \times 2000$\,pixel CCD in the blue arm and a Hamamatsu $2000 \times 4000$\,pixel CCD in the red arm. We observed with a D57 dichroic and a 600/4310 grism for the blue side and a 830/8460 grating for the red side, with respective dispersions of 0.43\,\AA/pixel and 1.02\,\AA/pixel. The approximate wavelength range covered by the blue arm was $3600$~--~$5300$\,\AA\ and by the red arm was $5700$~--~$7800$\,\AA. We used a slit width of $1.0$", which achieved a resolution of $\approx 1$\,\AA\ in the blue arm and $\approx 2$\,\AA\ in the red arm.

The first two observations, taken on 2018-05-22 and 2023-05-15, were published in \citet{Elms2023}. On 2018-05-22, one 3000\,s exposure was taken in the blue arm and three consecutive 1000\,s exposures were taken in the red arm. On 2023-05-15, two consecutive 2000\,s exposures were taken in the blue arm and four consecutive 1000\,s exposures were taken in the red arm. The third epoch observation was taken on 2023-06-25 and is published in this work for the first time. One 2700\,s exposure was taken in the blue arm and three consecutive 1200\,s exposures were taken in the red arm. Consecutive exposures are stacked in this work to create one blue and one red spectrum for each epoch. Spectroscopic coverage was achieved for all Balmer lines from H$\alpha$ to H$\zeta$.

\subsubsection{Magellan/MIKE spectroscopy}
\label{sec:MIKE}
Two spectra of WD\,J1653$-$1001 were taken with the Magellan Inamori Kyocera Echelle (MIKE) double echelle spectrograph \citep{Bernstein2003}. MIKE is mounted on the 6.5\,m Magellan 2 Clay Telescope located at Las Campanas Observatory, Chile.

One 1200\,s exposure was taken on 2023-05-28. WD\,J1653$-$1001 was observed again on 2023-09-21 for a total of 2400\,s in two exposures through a 1" wide and 5" long slit. We used the standard MIKE setup, with E2V $2000 \times 4000$\,pixel CCD detectors with the $2 \times 2$ binned readout mode and the standard gratings. The blue and red arms of the spectrograph were used simultaneously to provide full wavelength coverage from $3400$~--~$8950$\,\AA, therefore capturing the H$\alpha$ and H$\beta$ Balmer lines. Spectra were extracted, flat-fielded, wavelength-calibrated and averaged using the Carnegie Python pipeline \citep{Kelson2003}. 

\subsubsection{INT/IDS spectroscopy}
\label{sec:INT}
Spectroscopic observations of WD\,J1653$-$1001 were taken over two epochs using the Intermediate Dispersion Spectrograph (IDS) on the Cassegrain focus of the 2.5-m Isaac Newton Telescope (INT). The INT is located at the Observatorio del Roque de los Muchachos on La Palma, Spain. 

The first observation occurred on 2023-06-19, with 12 exposures each with exposure times of $\approx$\,1200\,s. Our setup utilized the EEV10 $4096 \times 2048$\,pixel CCD detector and the R600R grating. The spectra have a wavelength range $4380$~--~$6730$\,\AA\ therefore cover the H$\alpha$ and H$\beta$ Balmer lines. 

The second observation occurred on 2023-07-18, where nine $\approx$\,1200\,s exposures were taken. Our setup utilized the RED+2 $4096 \times 2048$\,pixel CCD detector and the R632V grating. The spectra have a wavelength range $4250$~--~$6730$\,\AA\ therefore also cover the H$\alpha$ and H$\beta$ Balmer lines. 

\subsubsection{Magellan/MagE spectroscopy}
\label{sec:MagE}
Three consecutive spectra with 1200\,s exposures were taken with the optical Magellan Echellette (MagE) Spectrograph \citep{Marshall2008} on 2024-07-11. MagE is mounted on the 6.5\,m Magellan 1 Baade Telescope located at Las Campanas Observatory, Chile. The observations were conducted with a 0.85$^{\prime\prime}$ slit, yielding wavelength coverage from $3700 - 9300$\,\AA\ at a resolving power of $R\approx4800$. Immediately following the science exposures, we obtained three ThAr arc frames to secure a precise wavelength calibration. Spectroscopic data reduction was carried out using \texttt{pypeit} \citep{Prochaska2020a, Prochaska2020b}. Additionally, we utilized the \texttt{merlin}\footnote{https://github.com/vedantchandra/merlin/tree/main} package, which offers an end-to-end reduction pipeline for MagE spectroscopic data built upon \texttt{pypeit} v1.15.0. The spectral wavelength coverage was adequate to capture the H$\alpha$ and H$\beta$ Balmer lines.

\subsubsection{MMT/Binospec spectroscopy}
\label{sec:Binospec}
WD\,J1653$-$1001 was observed on four consecutive nights, from 2025-05-19 to 2025-05-22, with the Binospec optical spectrograph \citep{Fabricant2019}. Binospec is mounted on the 6.5\,m Multiple Mirror Telescope (MMT) in Arizona, USA. We used the standard set-up with a $4000 \times 4000$\,pixel E2V CCD detector, $1.0$" wide long-slit mask, and LP3800 filter. We selected a wavelength coverage of $5740$~--~$7250$\,\AA\ to observe the H$\alpha$ Balmer region, with 1000 grating lines per mm and a dispersion of 0.36\,\AA/pixel.

A series of 20\,minute exposures were taken of WD\,J1653$-$1001. Six exposures were taken on 2025-05-19 and 2025-05-21, while five exposures were taken on 2025-05-22 due to increased overheads. The observing block on 2025-05-20 was terminated early due to the seeing increasing to $> 2$" in the last exposure, resulting in four 20\,minute exposures. We only use the first two exposures from 2025-05-20 for this work as they are comparable S/N as the other observations. The data were processed using the Center for Astrophysics Binospec pipeline (v1.99) and the individual exposures from each night were coadded to produce a single spectrum per night.

\subsection{Time-domain photometry}
\label{sec:Time-domain photometry}

\subsubsection{ZTF}
\label{sec:ZTF}
ZTF is a robotic time-domain survey which uses the 48-inch Schmidt Telescope at the Palomar Observatory in California, USA \citep{Masci2019}. In this work, we use publicly-available DR23 observations of WD\,J1653$-$1001 which were taken between 2018-03-17 and 2024-10-31 in two broadband filters: the green ($g$) filter with a bandpass between $4100-5500$\,\AA\ and a red ($r$) filter with a bandpass between $5500-7400$\,\AA\ \citep{Dekany2020}. Specifically for WD\,J1653$-$1001, the $g$-band has a baseline of 5.97\,yr and the $r$-band has a baseline of 6.35\,yr. The light curves were retrieved from the public NASA/IPAC Infrared Science Archive (\href{https://irsa.ipac.caltech.edu/frontpage/}{IRSA}). Exposure times of 30\,s were taken in all observations.

Since the publication of \citet{Elms2023} which used ZTF DR15, ZTF has collected an additional 2\,years worth of data. For WD\,J1653$-$1001, this equates to a data increase of 23 per cent in the red filter and 28 per cent in the green filter in DR23.

\subsubsection{ATLAS}
\label{sec:ATLAS}
ATLAS is an all-sky survey comprised of a network of four 0.5\,m telescopes, located in Hawaii, Chile and South Africa. Observations of WD\,J1653$-$1001 were taken between 2015-07-25 and 2025-03-19 in the main survey mode using 30\,s exposures and two non-standard broadband filters: a cyan ($c$) filter with a bandpass between $4200-6500$\,\AA\ and an orange ($o$) filter with a bandpass between $5600-8200$\,\AA\ \citep{Tonry2018}. Specifically for WD\,J1653$-$1001, the $c$-band has a baseline of 9.57\,yr and the $o$-band has a baseline of 9.65\,yr. We extracted publicly-available ATLAS light curves with the ATLAS forced photometry server \citep{Tonry2018, Smith2020, Shingles2021}, including a proper motion correction in the source position.

\subsubsection{Additional photometry}
\label{sec:Additional photometry}
We used the 1.23\,m telescope at the Calar Alto observatory to obtain time-series photometry with the ASI461MM Pro Complementary Metal-Oxide-Semiconductor (CMOS) camera. On the nights 2025-05-20 to 2025-05-23 and 2025-05-27 to 2025-05-30, observations in the Johnson $B$, $V$ and $R$ filters were taken across multiple wavebands by having multiple nights spaced close together in time. Furthermore, the transmission function of the $B$ filter is bluer than the filters used in ZTF and ATLAS, such that noticeable variability in this band could be an indicator of a variable emission line in Balmer lines bluer than H$\beta$. Exposure times ranged from 50-80\,s depending on the filter and the observing conditions to give a S/N ratio above 50 in each exposure. The general observing pattern was one hour of imaging in the $B$, $V$ then $R$ filter on each night. All observations were bias and dark current subtracted, flat field corrected and flux calibrated. 

We also obtained publicly-available photometric data for WD\,J1653$-$1001 from the All-Sky Automated Survey for Supernovae (ASAS-SN) Sky Patrol Photometry Database V2.0 \citep{Shappee2014, Hart2023}. ASAS-SN is a global network of 20 telescopes operated by the Las Cumbres Observatory \citep[LCO;][]{Brown2013}, distributed across Haleakala Observatory (Hawaii), Cerro Tololo International Observatory (Chile), McDonald Observatory (Texas) and the South African Astrophysical Observatory (South Africa). Observations of WD\,J1653$-$1001 began in 2013 using the legacy $V$-band filter. We do not use $V$-band photometric data in this work so only utilise the standard $g$-band data. ASAS-SN $g$-band observations of WD\,J1653$-$1001 range from 2018-01-23 to 2024-05-20, providing a baseline of 6.32\,yr. 

\section{Analysis}
\label{sec:Analysis}

\subsection{Photometric and spectroscopic variability}
\label{sec:Variability}
All observation time-stamps for WD\,J1653$-$1001 are converted from UTC to Barycentric Julian Date (BJD) in the Barycentric Dynamical Time (TDB) standard. For consistency, the time format used for all data is Barycentric Modified Julian Date (BMJD) minus 50\,000, which is BJD(TDB) $-$ 2\,450\,000.5. 

\subsubsection{Photometric variability}
\label{sec:Photometric variability of WDJ1653}

\begin{figure*}
\centering
\includegraphics[width=1.39\columnwidth]{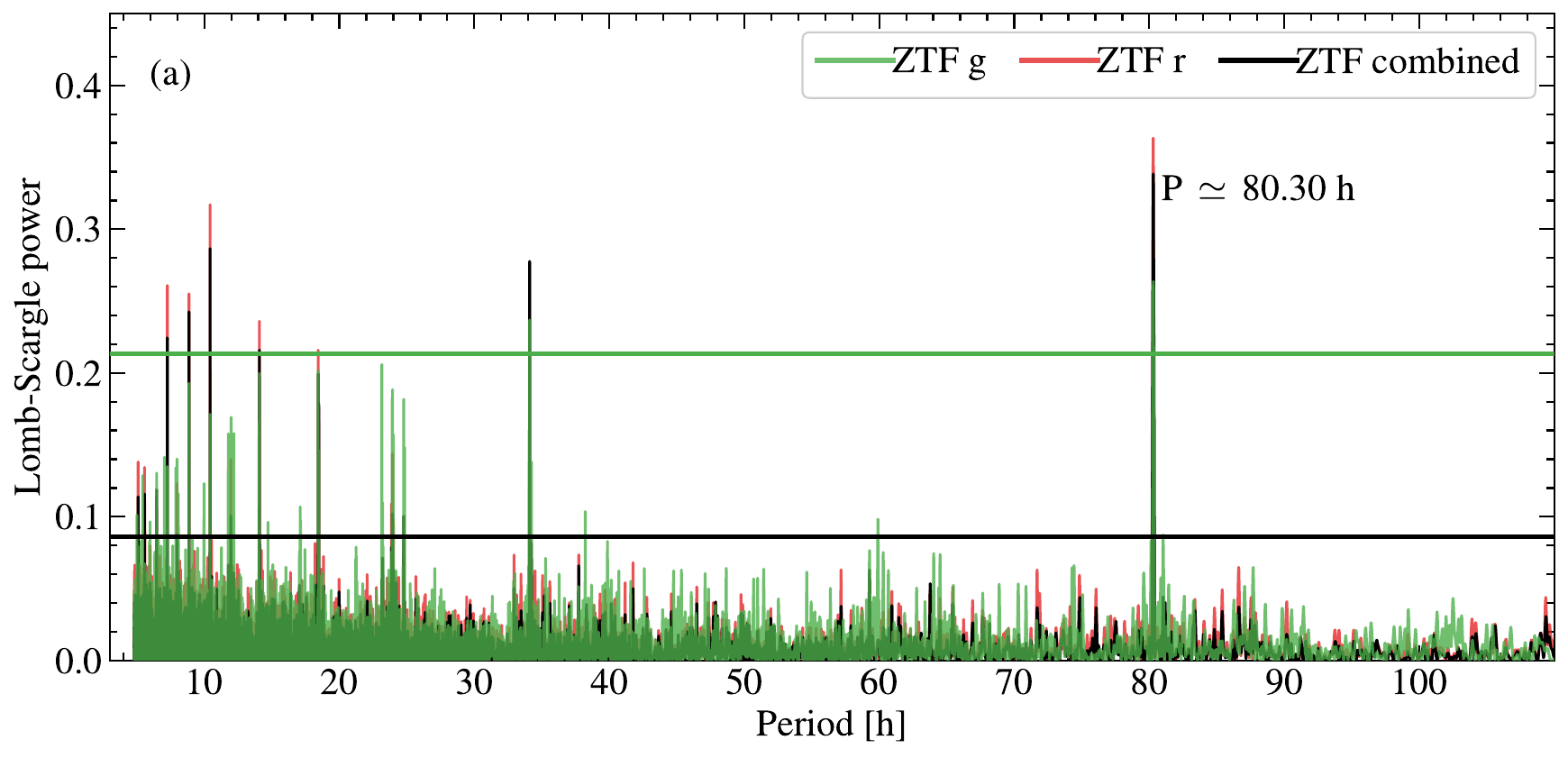}
\includegraphics[width=0.67\columnwidth]{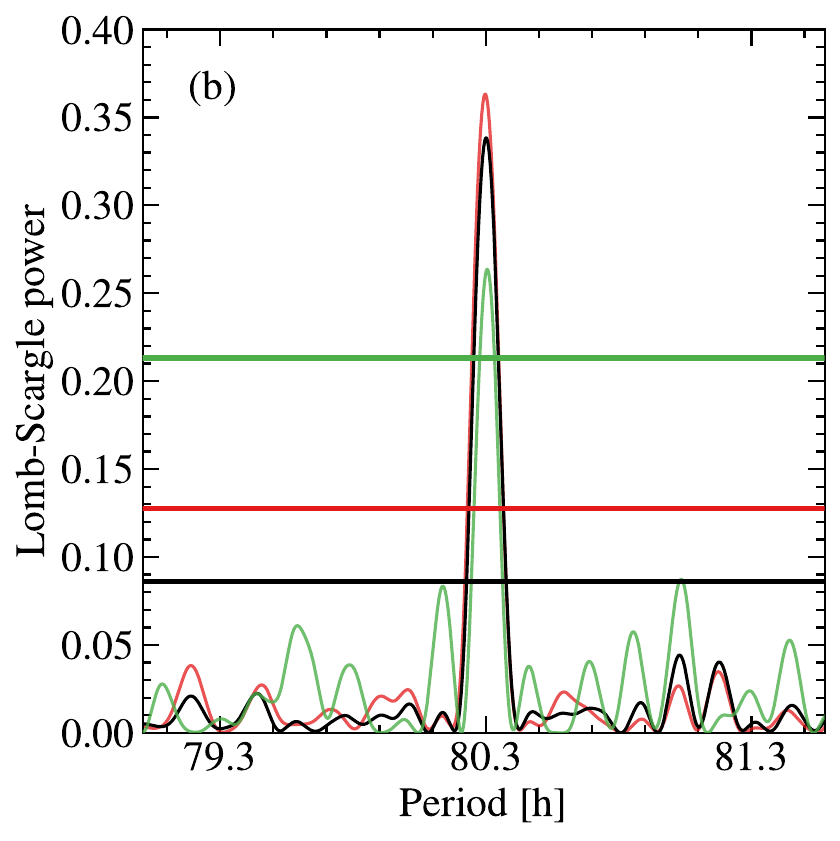}
\includegraphics[width=1.39\columnwidth]{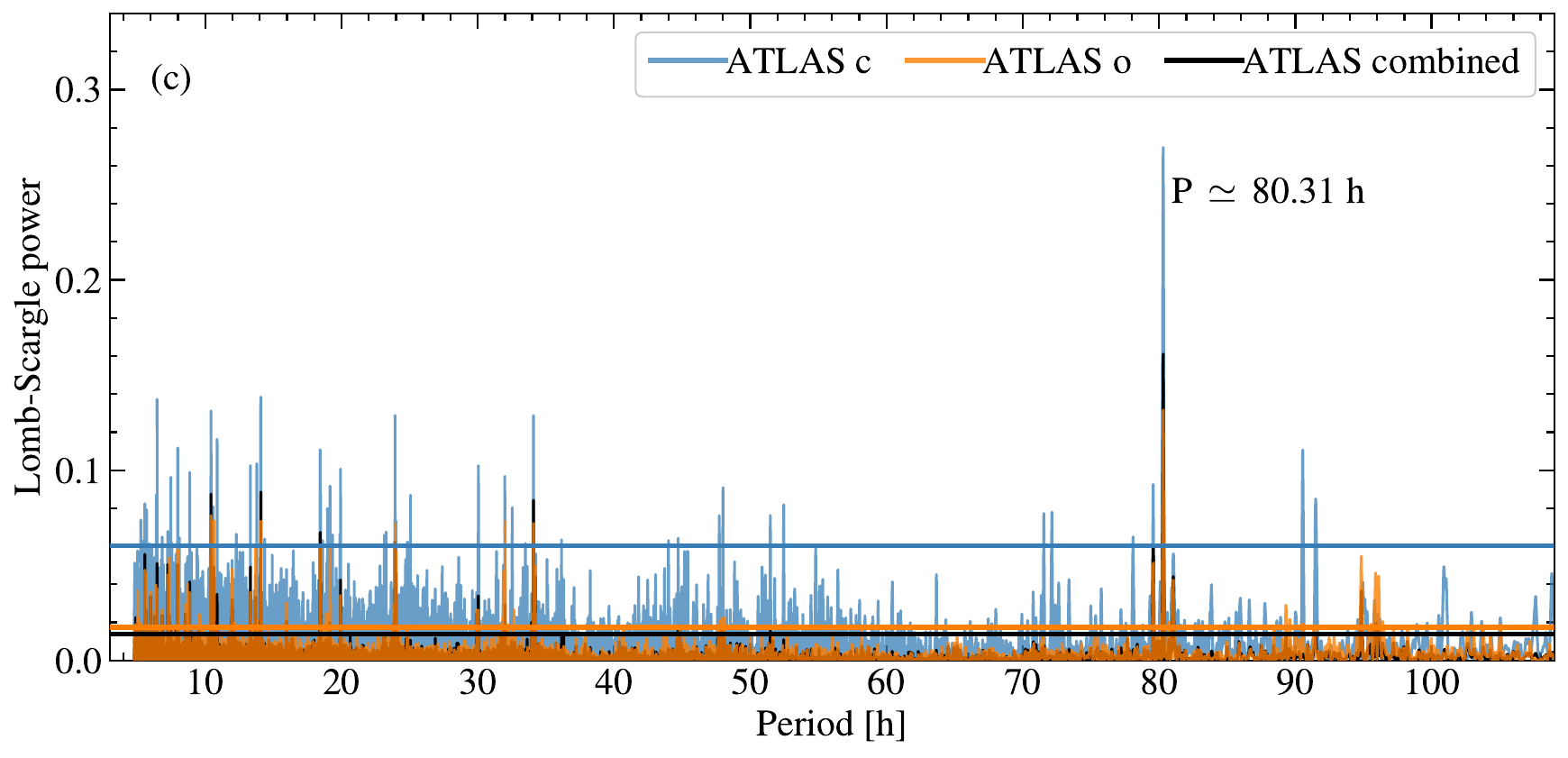}
\includegraphics[width=0.67\columnwidth]{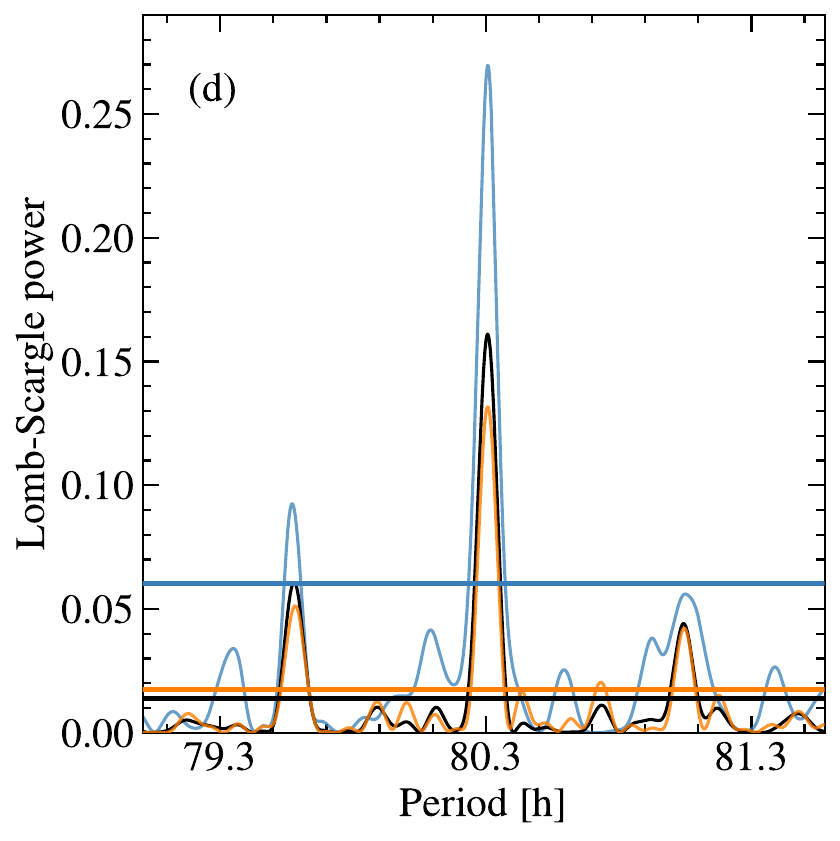}
    \caption{Power spectra computed from ZTF (a and b) and ATLAS (c and d) broad-band photometric data of WD\,J1653$-$1001. The strongest signal common to all power spectra corresponds to a period of $P \simeq 80.30 - 80.31$\,h, which is detected above a FAP of $5\sigma$ (solid horizontal lines). The DR23 ZTF $r$-band (red), $g$-band (green) and combined $g$- and $r$-band (black) data are shown in panels a and b, where b shows the power spectra zoomed in on the periodic signal at $P \simeq 80.30$\,h. The ATLAS $c$-band (blue), $o$-band (orange) and combined $c$- and $o$-band (black) data are shown in panels c and d, with d similarly zoomed in. The legends apply to both panels in each dataset.}
\label{fig:ZTF_ATLAS_power_spectra}
\end{figure*}

\begin{figure}
\centering
\raggedright
\includegraphics[width=0.981\columnwidth]{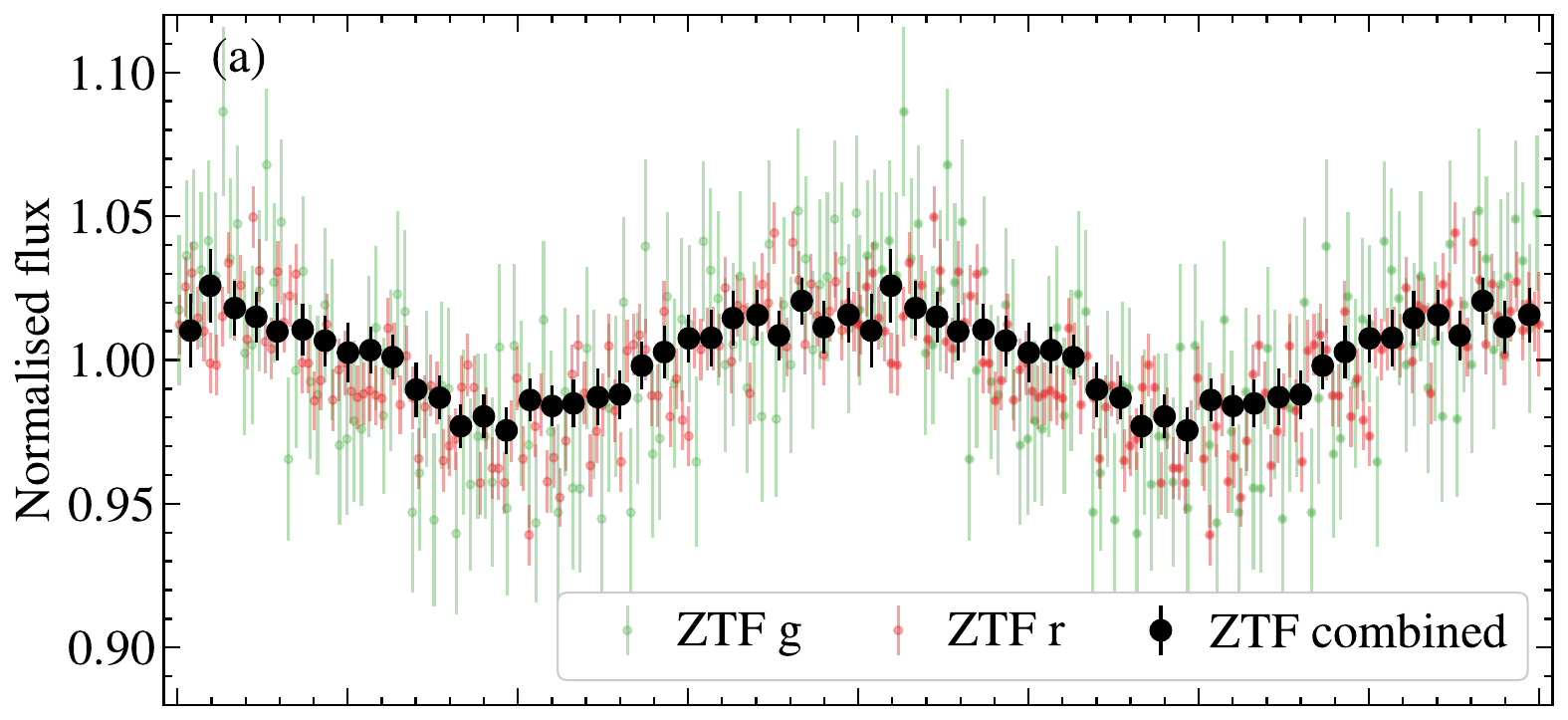}
\includegraphics[width=0.981\columnwidth]{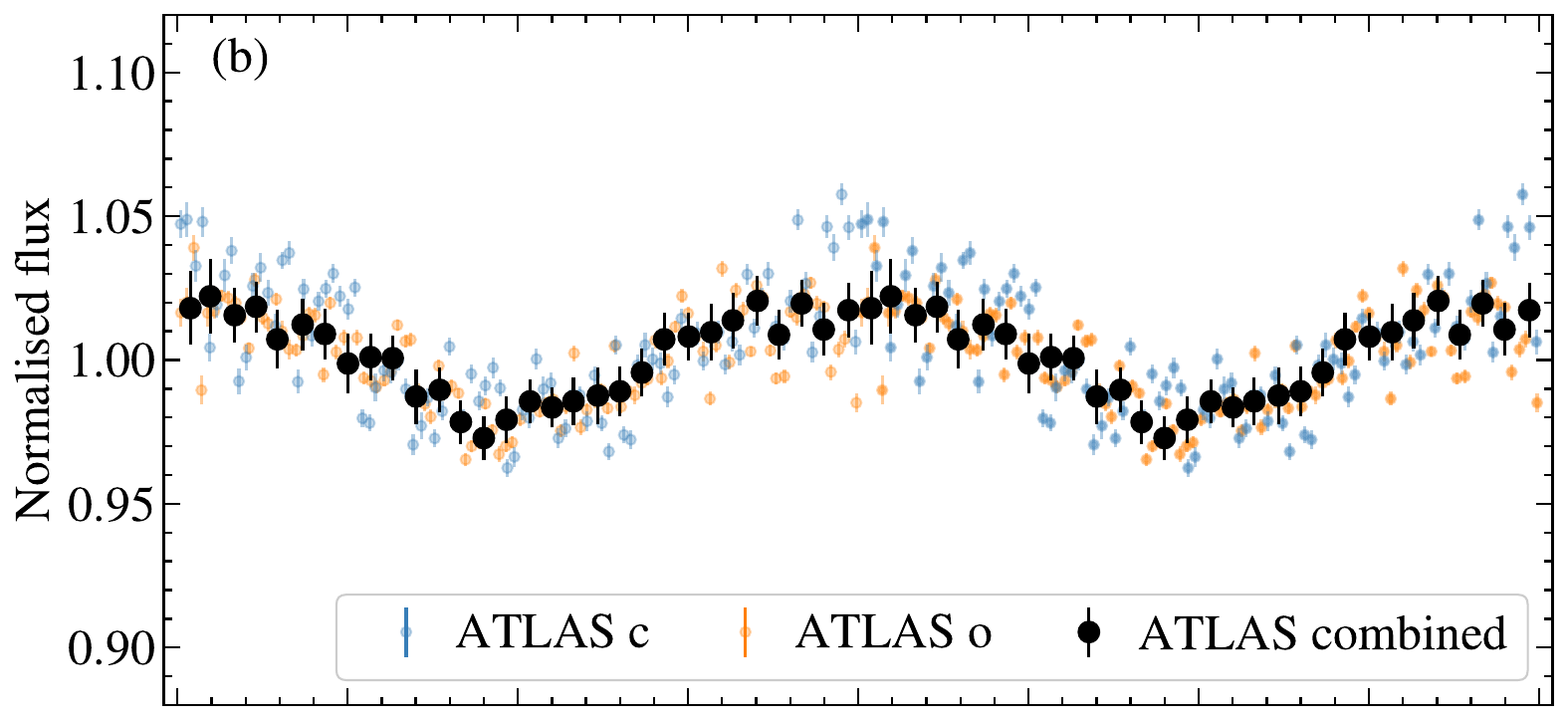}
\includegraphics[width=\columnwidth]{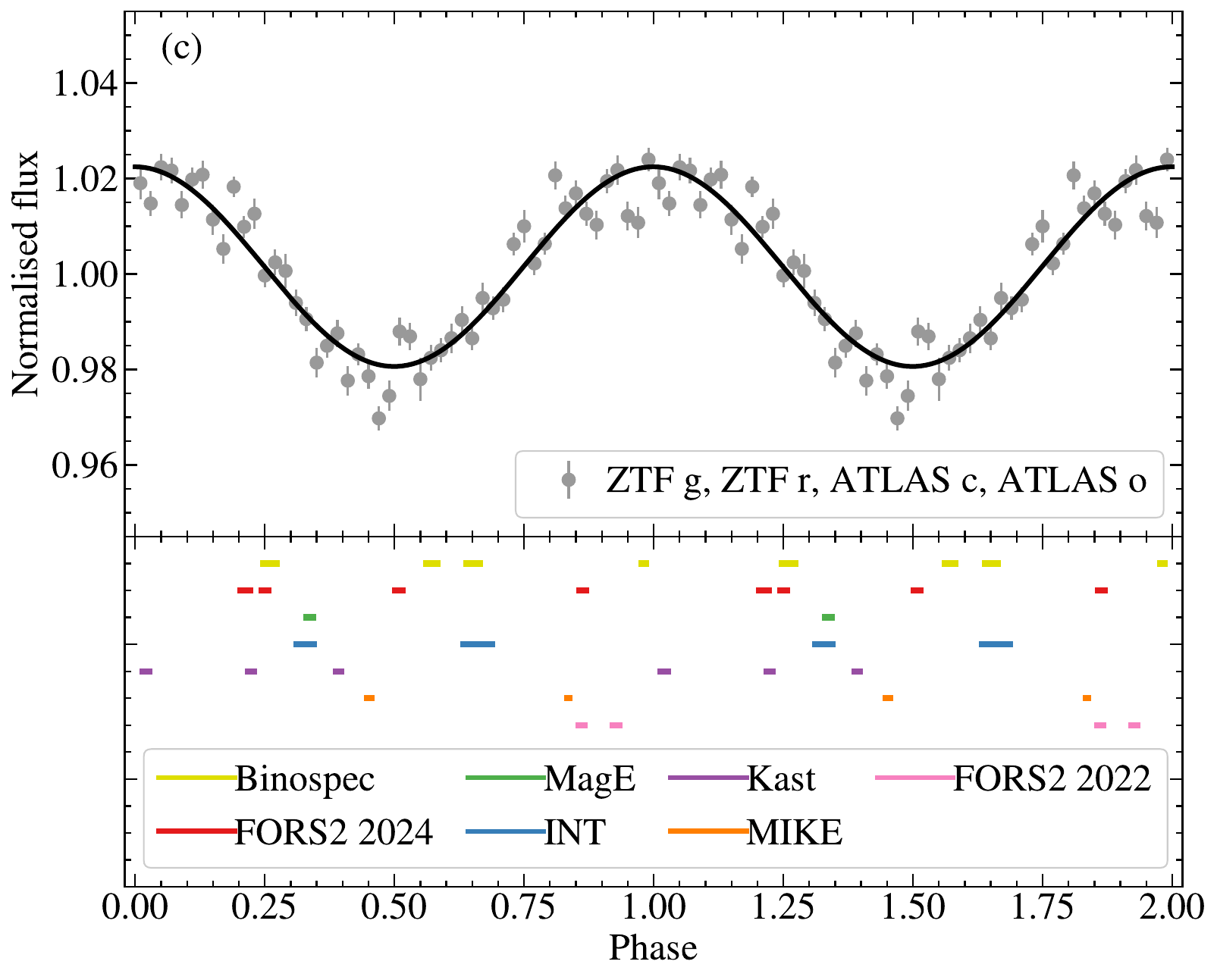}
    \caption{ZTF and ATLAS light curves of WD\,J1653$-$1001. (a) shows the ZTF $g$-band (green), $r$-band (red) and combined $g$- and $r$-band (black) light curves. (b) shows the ATLAS $c$-band (cyan), $o$-band (orange) and combined $c$- and $o$-band (black) light curves. (c) upper panel shows the combined $g$-, $r$-, $o$- and $c$-band (grey) light curve, fitted with a sinusoid (black solid line). ${\rm Phase} = 0$ corresponds to the photometric maximum at the ephemeris in Eq~\ref{ZTF_ATLAS_photometric_ephemeris}. All individual band light curves are binned into 150 data points and all combined light curves are binned into 30 data points. Each light curve is phase-folded onto a period of $80.3070$\,h and data are repeated over two phases for illustrative purposes. Error bars represent the 1$\sigma$ scatter in each bin. (c) lower panel shows the spectroscopic phase coverage achieved by observations from Binospec (yellow), FORS2 2024 (red), MagE (green), INT (blue), Kast (purple), MIKE (orange) and FORS2 2022 (pink).}
\label{fig:ZTF_ATLAS_lc_phase}
\end{figure}

\begin{table*}
    \centering
	\caption{Measured period values for WD\,J1653$-$1001 from ZTF and ATLAS individual $g$-, $r$-, $c$- and $o$-band datasets, in addition to the combined datasets. Three techniques were used: identification of the strongest signal in the power spectra; MCMC; and fitting a sine wave on the light curves. The best-fitting amplitudes for each light curve are also presented.}
	\label{tab:measured_periods}
	\begin{tabular}{llcccc}
		\hline
		\hline
		 Survey & Filter & Power spectra & MCMC & Fit sine curve & Amplitude\\
		 &  & [h] & [h] & [h] & [per cent]\\
		\hline
		\hline
		\multirow{3}{1cm}{ZTF} & $g$-band & 80.3049 & 80.3006 $\pm$ 0.0041 & 80.3054 $\pm$ 0.0099  & 3.08 $\pm$ 0.42\\
        & $r$-band & 80.2972 & 80.3000 $\pm$ 0.0042 & 80.2977 $\pm$ 0.0046 & 2.33 $\pm$ 0.14\\
        & Combined & 80.3015 & 80.2994 $\pm$ 0.0042 & 80.3012 $\pm$ 0.0043 & 2.39 $\pm$ 0.14\\
        \hline
        \multirow{3}{1cm}{ATLAS} & $c$-band & 80.3076 & 80.3066 $\pm$ 0.0029 & 80.3077 $\pm$ 0.0010 & 2.71 $\pm$ 0.06\\
        & $o$-band & 80.3070 & 80.3080 $\pm$ 0.0028 & 80.3071 $\pm$ 0.0009 & 1.87 $\pm$ 0.03\\
        & Combined & 80.3070 & 80.3077 $\pm$ 0.0021 & 80.3069 $\pm$ 0.0007 & 2.07 $\pm$ 0.03\\
        \hline
        ZTF and ATLAS & $g$-, $r$-, $c$-, $o$-band & 80.3070 & 80.3070 $\pm$ 0.0007 & 80.3073 $\pm$ 0.0007 & 2.09 $\pm$ 0.03 \\
        \hline
	\end{tabular}
\end{table*}

We make use of both ZTF and ATLAS time-domain photometry to analyse the variability of WD\,J1653$-$1001. The two surveys provide complementary datasets: ZTF delivers high-sensitivity, high-cadence measurements, while ATLAS provides a longer temporal baseline that extends the time coverage and improves sensitivity to long-term variability. Together, they enable a more complete characterisation of photometric behavior and help to confirm or reject potential aliases in the periodic signal.

The ZTF and ATLAS flux of WD\,J1653$-$1001 were calculated from the respective survey magnitude data, relative to the median magnitude in each band. We combined the individual datasets from each survey, i.e. the ZTF $g$- and $r$-bands and the ATLAS $c$- and $o$-bands, and weighted the contribution of the individual band points equally.

The ZTF light curves were cleaned by selecting only data flagged as reliable in the `ngoodobsrel' column for each filter, ensuring that no measurements associated with bad pixels were included. This initial selection yielded 195 data points in the $g$-band and 550 data points in the $r$-band. To mitigate potential biases introduced by deep-drilling epochs, which can affect signals with periods $>1$\,d, we excluded such epochs from the dataset. The $g$-band contained no deep-drilling epochs thus all 195 data points remained, whereas the $r$-band did include such epochs so 334 data points were retained after the removal. Finally, we excluded all anomalous measurements with normalised flux values outside the range $0.89 < F_\mathrm{norm} < 1.09$. The final cleaned dataset consists of 188 data points in the $g$-band, 334 data points in the $r$-band and a total of 522 data points in the combined light curve. 

With the cleaned data, we computed Lomb-Scargle periodograms \citep{Lomb1976, Scargle1982} using the \texttt{python} package \texttt{astropy.timeseries} \citep{Astropy2013, Astropy2018, Astropy2022} for the individual $g$-band and $r$-band light curves in addition to the combined $g$- and $r$-band light curve (Figure~\ref{fig:ZTF_ATLAS_power_spectra}a and b). We analysed the $g$-band, $r$-band and combined power spectra to identify the strongest unique signals (separated by $\geq 1$\,h) at periods $< 28$\,d. We selected the top five detections in order of power then imposed a False Alarm Probability (FAP) threshold of $5\sigma$ on all signals. This analysis reveals two signals in the $g$-band power spectrum, with the most significant being 80.3049\,h and the second being 34.0984\,h. In order of power, the $r$-band power spectrum exhibits signals at: 80.3007\,h, 10.4274\,h, 34.0989\,h, 7.2632\,h and 8.8672\,h. Similarly, the combined power spectrum yields: 80.3015\,h, 10.4276\,h, 34.0989\,h, 8.8673\,h and 7.2632\,h. The periodic signal at $P \simeq 80.30$\,h is the only signal detected consistently, robustly and at the highest power across all three ZTF datasets. This coherence across datasets, together with its dominance in power, identifies $\simeq 80.30$\,h as the most likely true periodic signal of WD\,J1653$-$1001. The signal at $P = 34.10$\,h appears consistently in each power spectrum (at approximately the same power in the $r$-band and combined power spectra hence the $r$-band signal is concealed in Figure~\ref{fig:ZTF_ATLAS_power_spectra}a), however it is always weaker than the $80.30$\,h signal and does not dominate any dataset. Three additional signals at varying power levels are present in only two out of the three power spectra. Because these signals are not detected coherently across all ZTF datasets and are consistently weaker than the $80.30$\,h signal, we do not interpret them as true periodic signals but instead attribute them to sampling effects, aliases or band-dependent systematics. Table~\ref{tab:measured_periods} displays the strongest periodic signal measured in all three ZTF power spectra.

The broadband time-series forced photometry from ATLAS was cleaned similarly to the ZTF data before analysis. We imposed our own quality cuts on the data, selecting only reliable data which have: a magnitude error $< 1.0$\,mag; flux error $< 1000$\,$\mu$Jy; and a reduced $\chi^2 / N$ of the point‐spread‐function (PSF) fit for that forced photometry measurement of $<4.0$. This selection yielded 881 data points in the $c$-band and 3445 data points in the $o$-band. No epochs of this ATLAS data contained deep-drilling so all data points remained from this cut. Finally, we excluded all anomalous measurements with normalised flux values outside the range $0.89 < F_\mathrm{norm} < 1.09$, which left 751 data points in the $c$-band, 2815 data points in the $o$-band and a total of 3566 data points in the combined light curve.

As with the ZTF data, we created Lomb-Scargle periodograms of the individual ATLAS filter and combined data sets (Figure~\ref{fig:ZTF_ATLAS_power_spectra}c and d), identified the strongest unique signals at periods $< 28$\,d, selected the top five detections in order of power and imposed a FAP threshold of $5\sigma$ on all signals. This yields periodic signals in the $c$-band of: 80.3081\,h, 14.0854\,h, 6.4700\,h, 10.4277\,h and 23.9331\,h. In the $o$-band, the periodic signals are: 80.3069\,h, 10.4275\,h, 14.0853\,h, 32.0121\,h and 34.0938\,h. The combined dataset yields periodic signals of: 80.3069\,h, 14.0853\,h, 10.4276\,h, 34.0938\,h and 18.4377\,h. The periodic signal at $P \simeq 80.31$\,h has the highest power across all three ATLAS datasets, consistent with the strongest signals in the three ZTF power spectra. The signals at $P = 14.09$\,h and  $P = 10.43$\,h appear in each ATLAS power spectrum but are significantly weaker than the $P = 80.31$\,h signal and never dominate any dataset. One additional signal is present in two out of the three power spectra, at varying power levels, but due to its lack of coherence across all ATLAS datasets and its comparatively low power, we do not consider it a true periodic signal. Table~\ref{tab:measured_periods} displays the strongest periodic signal measured in all three power spectra.

In the six power spectra from ZTF and ATLAS, only the periodic signal at $P = 80.3$\,h is consistent across all datasets. Furthermore, it is the dominant signal in all six power spectra. While several weaker periodic signals appear in subsets of the ZTF and ATLAS data, none are coherent across all six power spectra nor do they approach the significance of the $80.3$\,h signal. 

As a final test, we combined the ZTF and ATLAS individual $g$-, $r$-, $c$- and $o$-band datasets and created a Lomb-Scargle periodogram, of which the strongest signal is $P = 80.3070$\,h. This further confirms the $80.3$\,h period as the true periodic signal of the white dwarf.

We performed least-squares fits to the ZTF and ATLAS light curves using the sinusoidal signal 
\begin{equation}
 \Delta \mathrm{flux} = A\mathrm{sin}(2\pi (t - T_0) / P) + c ,
 \label{sine_function}
\end{equation}
{\noindent}where $A$ is the amplitude, $t$ is the observation time of each measurement, $T_0$ is the epoch time at phase zero, $P$ is the period, and $c$ is a flux offset. The fitting was performed using the non-linear least-squares trust region reflective \citep[\texttt{trf};][]{Byrd1987} algorithm implemented in \texttt{scipy optimize} to determine the best-fitting $A$, $P$ and $c$. No parameter bounds were imposed, ensuring an unrestricted exploration of the parameter space. The best-fitting periods and amplitudes for each light curve are reported in Table~\ref{tab:measured_periods}. 

Finally, we estimated the period uncertainty in each filter using a Markov chain Monte Carlo (MCMC). Using the Python package \texttt{emcee} \citep{ForemanMackey2013}, we fit a sinusoid to all available data for a given survey and filter, with four free parameters: period, amplitude, phase offset and mean flux. In addition, we include a nuisance, or jitter, parameter to account for additional variability in the data not captured by the reported photometric uncertainty. We adopt uniform (flat) priors on all five independent parameters, with the period confined to the range 80.2--80.4\,h. We find well-converged, Gaussian posteriors for all four parameters in all four fits, with standard deviations on the period ranging from 0.0028--0.0042\,h. In Table~\ref{tab:measured_periods} we show the median periods, as determined from the posterior distributions, with the associated uncertainty defined as the standard deviation of the posterior. We find that all four filters yield statistically-consistent periods.

In addition to fitting the individual light curves, we also perform three simultaneous fits. In the first, we fit the ZTF $g$- and $r$-band light curves, each fitted with an 4-parameter sinusoid. The parameters of each sinusoid are fully independent, with the exception of the period which is common to both sinusoids, yielding a 7-parameter fit. This procedure yields a best-fit period of $P$\,=\,80.2994\,$\pm$\,0.0042\,h. We repeat this procedure for the two ATLAS light curves ($c$ and $o$), finding a best-fit period of $P$\,=\,80.3077\,$\pm$\,0.0021\,h. The smaller uncertainty on the ATLAS period originates from the longer baseline and larger dataset of the ATLAS data compared to ZTF. Finally, we perform a 13-parameter fit, for four independent sinusoids (one for each $g$-, $r$-, $c$- and $o$-band) with a common period. This procedure yields a best-fitting period of $P$\,=\,80.3070\,$\pm$\,0.0007\,h.

Figure~\ref{fig:ZTF_ATLAS_lc_phase} shows the individual and combined light curves for the ZTF filters in (a) and for the ATLAS filters in (b). The upper panel of Figure~\ref{fig:ZTF_ATLAS_lc_phase}(c) shows the light curve from the combined $g$-, $r$-, $c$- and $o$-band datasets with a fitted sinusoidal. The photometric ephemeris, quoted at the center of the time baseline of the dataset and where the photometric maximum is phase zero, is 
\begin{equation}
 \rm BMJD-50\,000 = 8991.85145(41) + 3.34613(03) \textit{E} ,\label{ZTF_ATLAS_photometric_ephemeris}
\end{equation}
\noindent where $E$ is the cycle count and $3.34613 \pm 0.00003$\,d corresponds to the best-fitting period from the simultaneous $g$-, $r$-, $c$- and $o$-band MCMC. We use Eq.~\ref{ZTF_ATLAS_photometric_ephemeris} to phase the spectroscopic observations of WD\,J1653$-$1001 (Section~\ref{sec:Spectroscopic variability}). 

\begin{figure}
\centering
\includegraphics[width=\columnwidth]{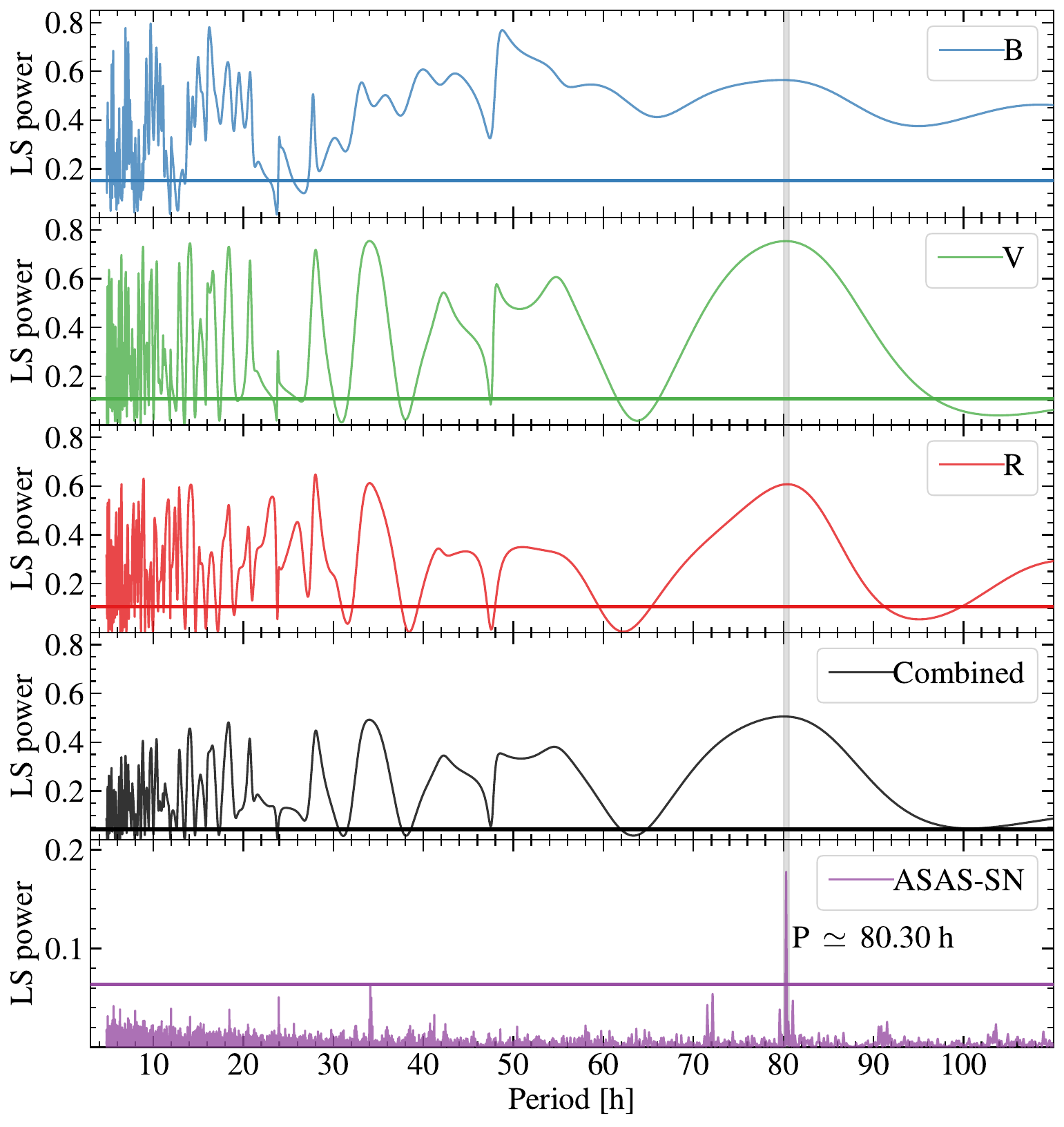}
    \caption{Power spectra computed from Calar Alto and ASAS-SN photometric data of WD\,J1653$-$1001. A consistent signal of $P \simeq 80.3$\,h (grey vertical line) is detected above a FAP of $5\sigma$ (solid horizontal lines) in the $B$ (blue), $V$ (green), $R$ (red) and combined (black) filter data from Calar Alto, and also in the ASAS-SN (purple) data.}
\label{fig:CA_ASAS_power_spectra}
\end{figure}

We computed Lomb-Scargle periodograms of the Calar Alto time-series photometry (Section~\ref{sec:Additional photometry}) in order to independently verify the periodic signal identified in the ZTF and ATLAS data. The Lomb-Scargle periodograms constructed from the individual $B$, $V$ and $R$ filter data, as well as from the combined dataset, are presented in Fig.~\ref{fig:CA_ASAS_power_spectra}. The strongest and most consistent signal in the $V$, $R$ and combined filter power spectra occurs at a period of $P \approx 80$\,h. This signal is also present in the $B$ filter periodogram, though at a lower power, and hints that there may be some faint variability across Balmer features bluer than H$\beta$. A least-squares sinusoidal fit of the combined filter data has a best-fitting period of $80.0589 \pm 0.3134$\,h and an MCMC analysis reveals $80.40~\pm~0.33$\,h as the best-fitting period. 

We also analysed additional photometric data of WD\,J1653$-$1001 from the ASAS-SN $g$-band (Section~\ref{sec:Additional photometry}), similarly to ZTF and ATLAS. The light curve was cleaned by selecting only data with a `good' quality flag, errors $< 1.0$, and with normalised flux values $0.89 < F_\mathrm{norm} < 1.09$. This yields 720 data points. A Lomb-Scargle periodogram identifies the strongest unique signal at $P = 80.3038$\,h (Fig.~\ref{fig:CA_ASAS_power_spectra}) and a least-squares sinusoidal fit has a best-fitting period of $80.3038 \pm 0.0056$\,h and amplitude of $2.26 \pm 0.20$ per cent. An MCMC analysis reveals $80.3087 \pm 0.0047$\,h as the best-fitting period. 

We investigated the possibility that the period of variability measured with ZTF and ATLAS is an alias. Some DAHe white dwarfs in \citet{Manser2023, Reding2023} are thought to have two spots/regions beneath the chromosphere, causing the true period of variability to be $2P$. For WD\,J1653$-$1001, phase-folding the light curves on $2P = 160.6$\,h yields two maxima and minima within one phase. However, the ZTF and ATLAS power spectra signals at $2P$ are all below the $5\sigma$ FAP levels. As there is no significant signal present at $2P$, and spectropolarimetry measurements suggest we only observe one pole of the star (Section~\ref{sec:detection of a magnetic field}), we discount the true period of variability to be $2P$ with the current data. We also investigated whether the true period of variability could be $P/2 = 40.15$\,h, of which the signals are a mixture of under or marginally over the FAP of the ZTF and ATLAS bands. However, the signals at $P/2$ are all within the repeated alias pattern of the power spectra and cannot be reliably distinguished as a rotation period.

\subsubsection{Spectroscopic variability}
\label{sec:Spectroscopic variability} 
WD\,J1653$-$1001 was observed spectroscopically 20 times from 2018 to 2025 with six different instruments (see Table~\ref{tab:observations} for details). These observations consist of a mixture of single and multiple exposures. The observations provide $24.9$ per cent spectroscopic phase coverage of WD\,J1653$-$1001, which is a significant improvement from the $2.4$ per cent in \citet{Elms2023}. The spectroscopic phase coverage of each individual telescope/instrument is visualised in the lower panel of Figure~\ref{fig:ZTF_ATLAS_lc_phase}(c). 

\begin{figure}
	\includegraphics[width=1\columnwidth]{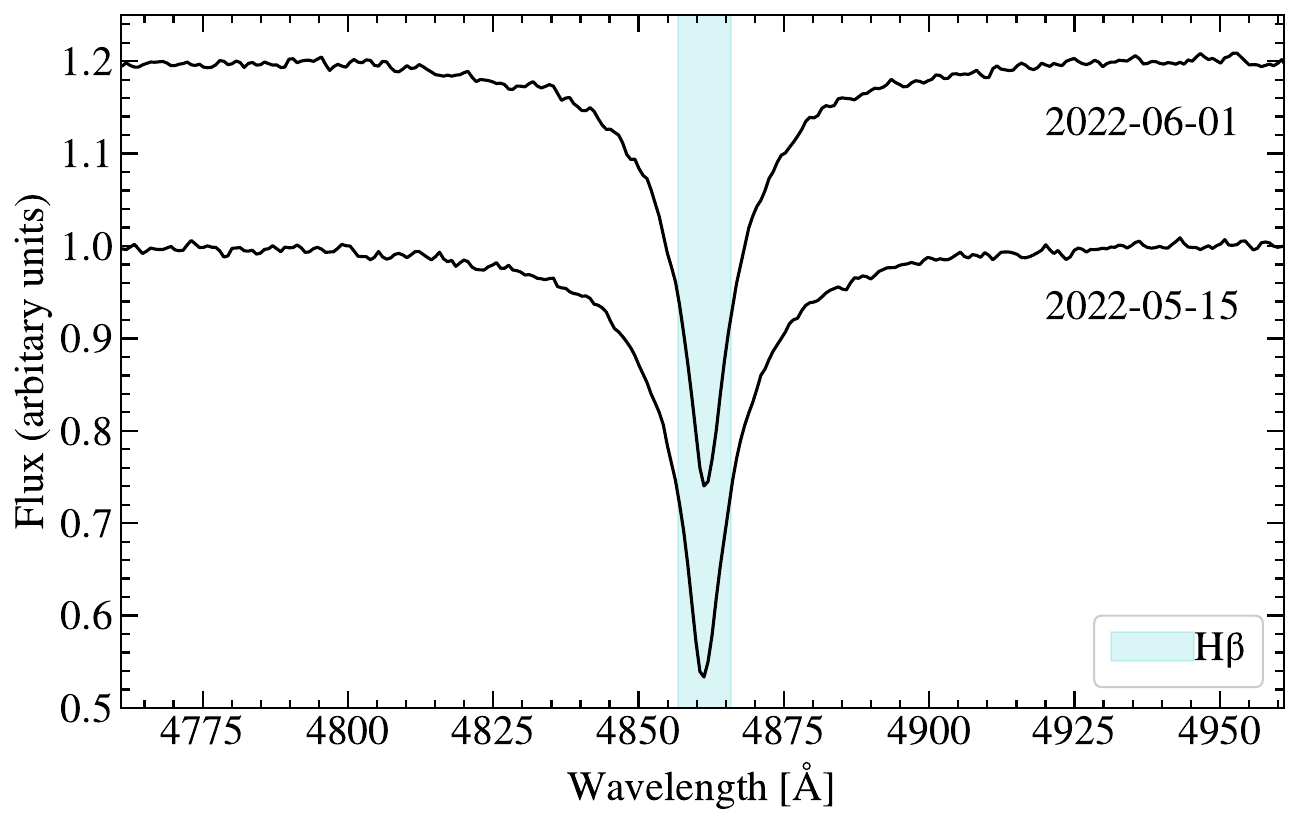}
    \caption{Two spectra of WD\,J1653$-$1001 taken with the \textbf{FORS2} spectrograph on 2022-05-15 and 2022-06-01 around the H$\beta$ Balmer line region. The observation UT dates are shown on the right of the plot. Spectra are convolved with a Gaussian with a FWHM of 1\,\AA\ and offset vertically for clarity.}
    \label{fig:FORS2_2022_spectra}
\end{figure}

\begin{figure}
	\includegraphics[width=1\columnwidth]{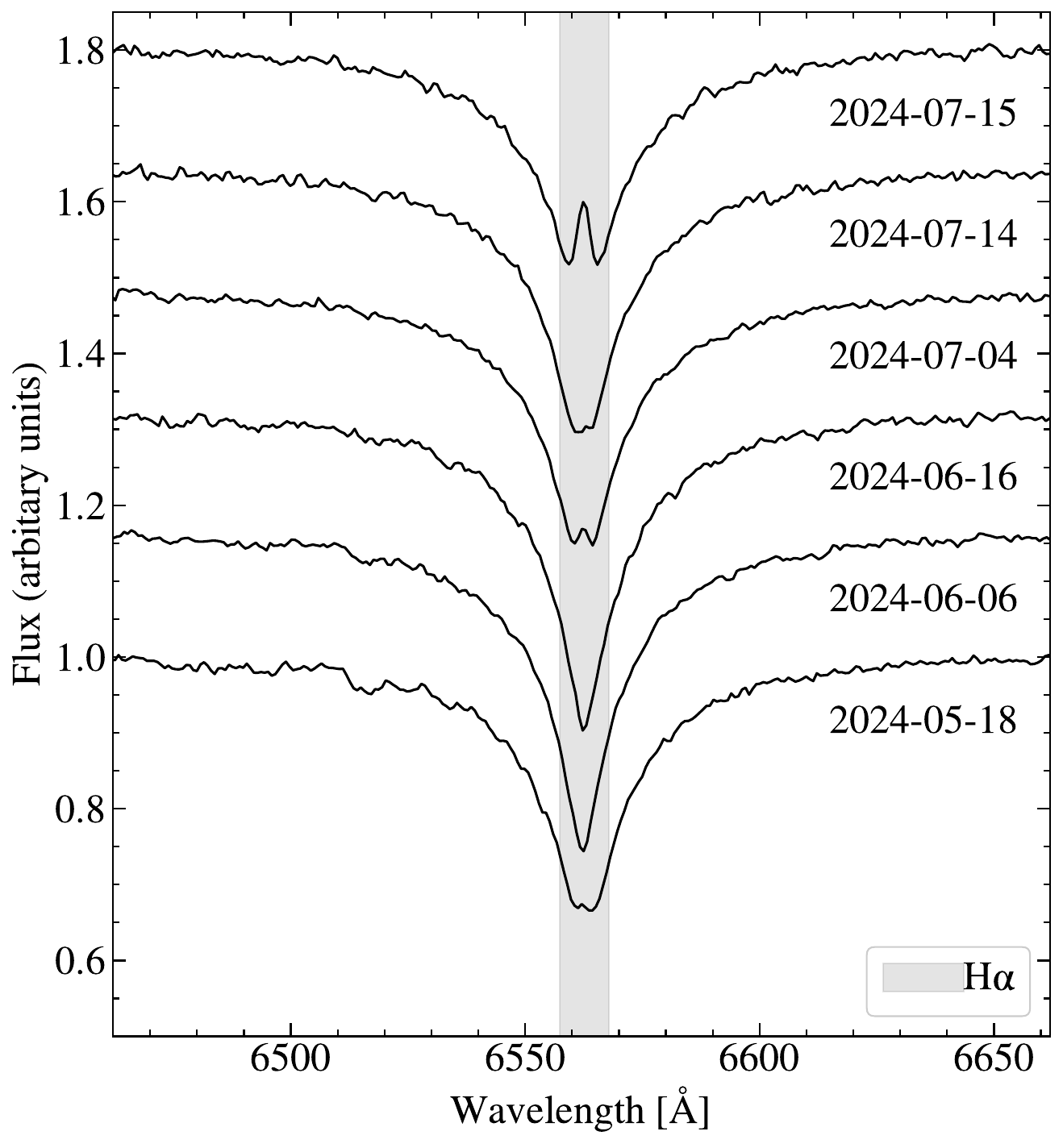}
    \caption{Six spectra of WD\,J1653$-$1001 taken with the \textbf{FORS2} spectrograph between 2024 May-July around the H$\alpha$ Balmer line region. The observation UT dates are shown on the right of the plot. Spectra are convolved with a Gaussian with a FWHM of 1\,\AA\ and offset vertically for clarity.}
    \label{fig:FORS2_2024_spectra}
\end{figure}

The observations taken with the FORS2 spectrograph on the VLT (Section~\ref{sec:FORS2}) are shown in Figures~\ref{fig:FORS2_2022_spectra}~and~\ref{fig:FORS2_2024_spectra}. The two exposures in Figure~\ref{fig:FORS2_2022_spectra} from 2022 show the stacked spectra from the four individual exposures each night. Spectra are centered around the H$\beta$ Balmer line region and do not evidence a visible emission spike in the line cores. Due to the relatively long variability period of WD\,J1653$-$1001, these observations caught this star at times of weak emission. Conversely, the six stacked spectra in Figure~\ref{fig:FORS2_2024_spectra} are from the eight 330\,s exposures taken on six nights from May to July 2024. These are centered around the H$\alpha$ Balmer line region and core emission variability is evident throughout the exposures. These data were reduced using daytime calibrations and not corrected for barycentric velocity, in order to avoid spurious polarisation signals that would result from the pipeline's frame-by-frame wavelength corrections, resulting in a small apparent velocity offset which also reflects instrumental flexure between calibration and science exposures. As this offset is not an astrophysical velocity shift, we corrected it by visually aligning the observed spectral features with their expected rest wavelengths.

\begin{figure*}
	\includegraphics[width=2\columnwidth]{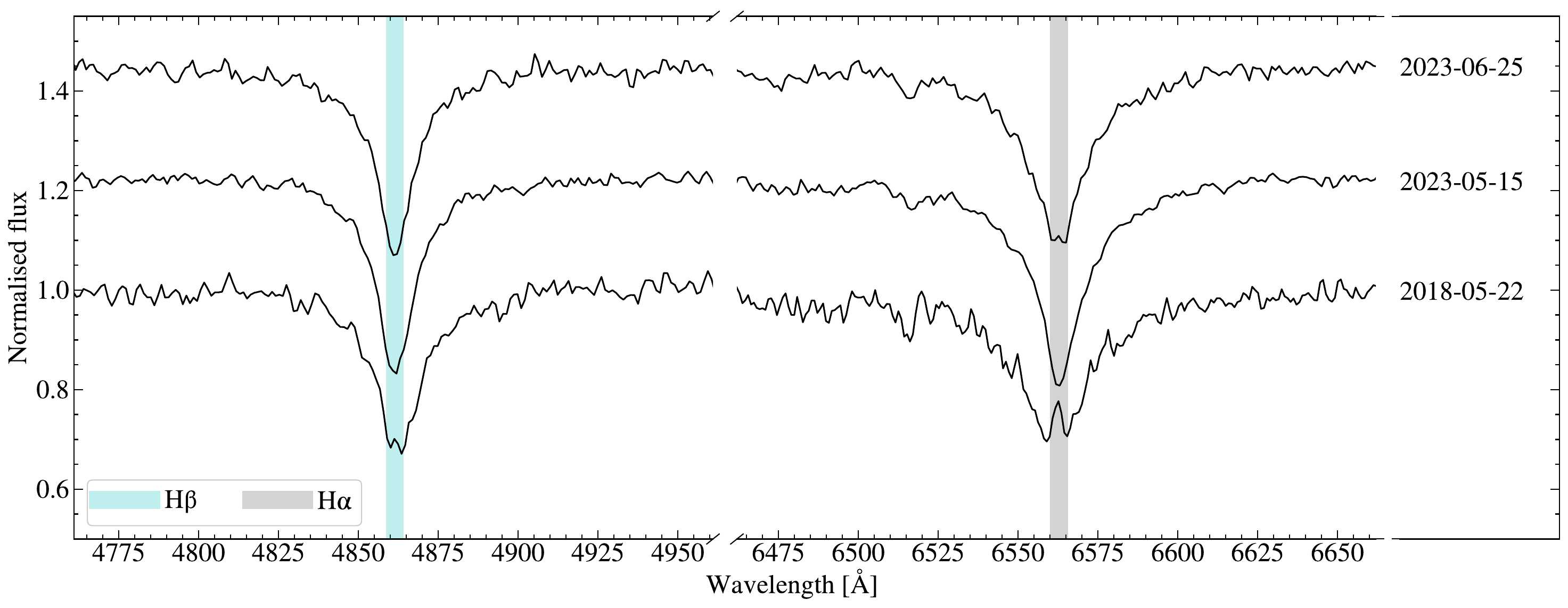}
    \caption{Three spectra of WD\,J1653$-$1001 taken with the \textbf{Kast} spectrograph from 2018 to 2023 around the H$\alpha$ and H$\beta$ Balmer line regions. The observation UT dates are shown on the right of the plot. Spectra are convolved with a Gaussian with a FWHM of 1\,\AA\ and offset vertically for clarity.}
    \label{fig:Kast_spectra}
\end{figure*}

The three spectra of WD\,J1653$-$1001 taken with the Kast instrument (Section~\ref{sec:Kast}) are shown in Figure~\ref{fig:Kast_spectra}, which consists of the individual exposures taken in the blue arm on 2018-05-22 and 2023-06-25, and the stacked spectra from the other exposures (Table~\ref{tab:observations}). Coverage of the H$\alpha$ and H$\beta$ Balmer line regions were achieved. The emission core variability is clearly visible between epochs. Small wavelength offsets are present in each spectrum that are consistent with typical Kast instrumental and wavelength calibration uncertainties. We corrected these offsets to visually align the observed spectral features with their expected rest wavelengths and do not interpret these offsets as a physical radial velocity.

\begin{figure*}
	\includegraphics[width=2\columnwidth]{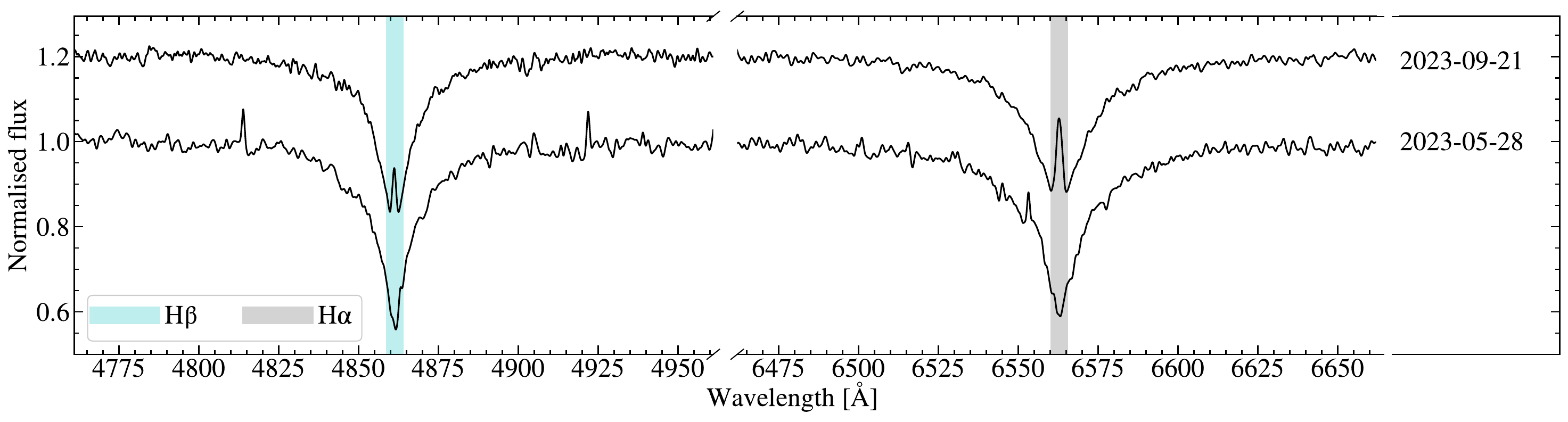}
    \caption{Two spectra of WD\,J1653$-$1001 taken with the \textbf{MIKE} spectrograph on 2023-05-28 and 2023-09-21 around the H$\alpha$ and H$\beta$ Balmer line regions. The observation UT date is shown on the right of the plot. The spectra are convolved with a Gaussian with a FWHM of 1\,\AA.}
    \label{fig:MIKE_spectra}
\end{figure*}

Figure~\ref{fig:MIKE_spectra} shows two spectra of WD\,J1653$-$1001 taken with the MIKE instrument (Section~\ref{sec:MIKE}) on different epochs. Coverage of the H$\alpha$ and H$\beta$ Balmer line regions were achieved. These observations caught this star at times of weak and strong emission.

\begin{figure*}
	\includegraphics[width=2\columnwidth]{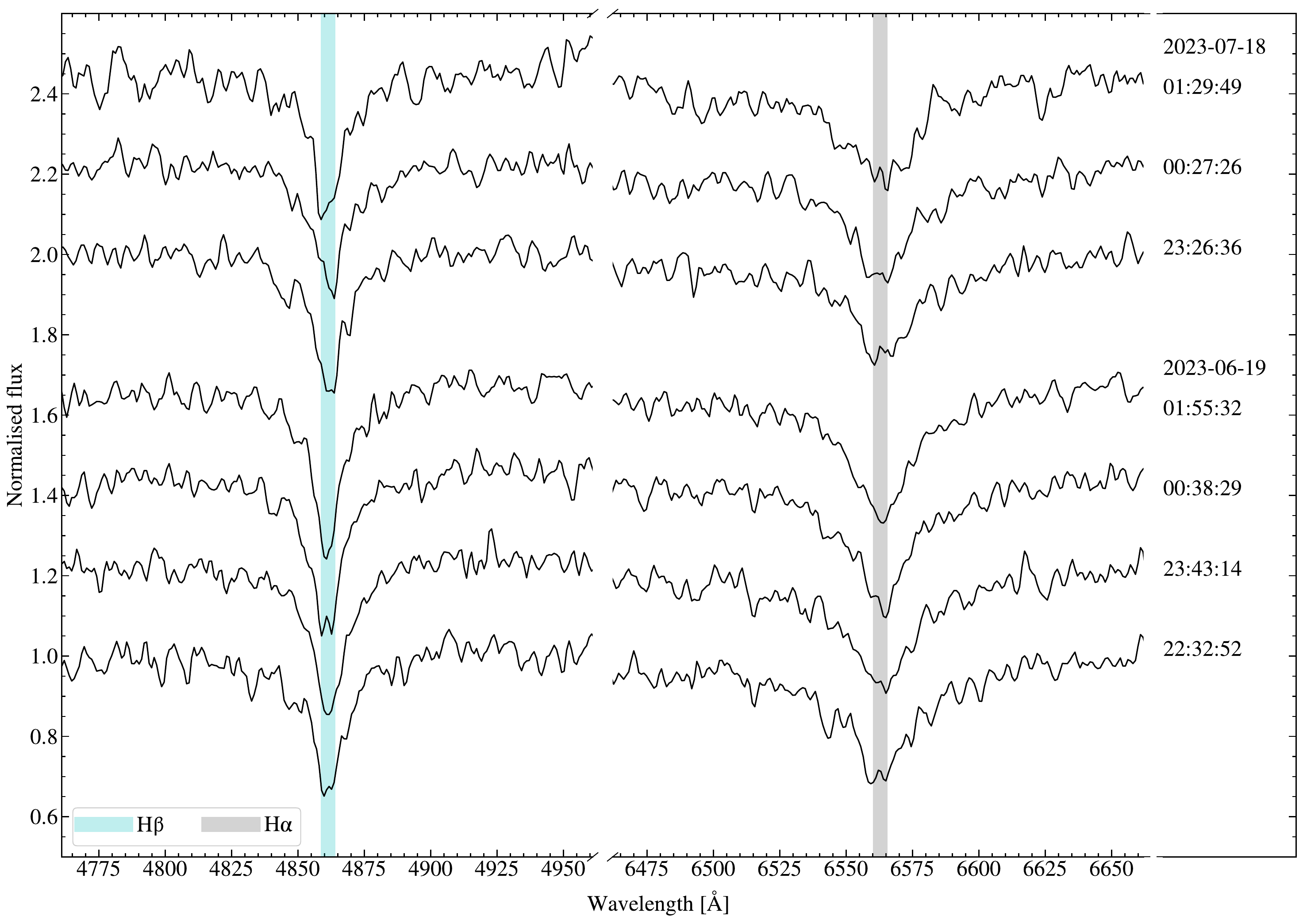}
    \caption{Stacked spectra of WD\,J1653$-$1001 taken with the \textbf{IDS} spectrograph between 2023 June-July around the H$\alpha$ and H$\beta$ Balmer line regions. The observation UT dates and times are shown on the right of the plot, where the date refers to at mid-exposure. Spectra are convolved with a Gaussian with a FWHM of 2\,\AA\ and offset vertically for clarity.}
    \label{fig:IDS_spectra}
\end{figure*}

The H$\alpha$ and H$\beta$ Balmer line regions of WD\,J1653$-$1001 were observed with $\approx 20$\,minute exposures with the IDS instrument on the INT (Section~\ref{sec:INT}) therefore, due to the relatively long spin period of this star, exposures were stacked to emphasise the Balmer line core emission. Figure~\ref{fig:IDS_spectra} shows the stacked IDS spectra. All 12 exposures taken in June were used in the stack, whereas 8 of 9 July exposures were used; the final July exposure was excluded due to its shorter exposure time and significantly lower S/N. Spectra were flux-scaled to the highest S/N spectrum in each dataset, then grouped by hour, with the mean of each group used as the stacked spectrum. The time at mid-exposure of each stacked spectrum is shown on the right-hand side of each spectrum. Balmer line core emission is less pronounced in these spectra than other observations of WD\,J1653$-$1001, nevertheless variability between exposures is still visible. 

\begin{figure*}
	\includegraphics[width=2\columnwidth]{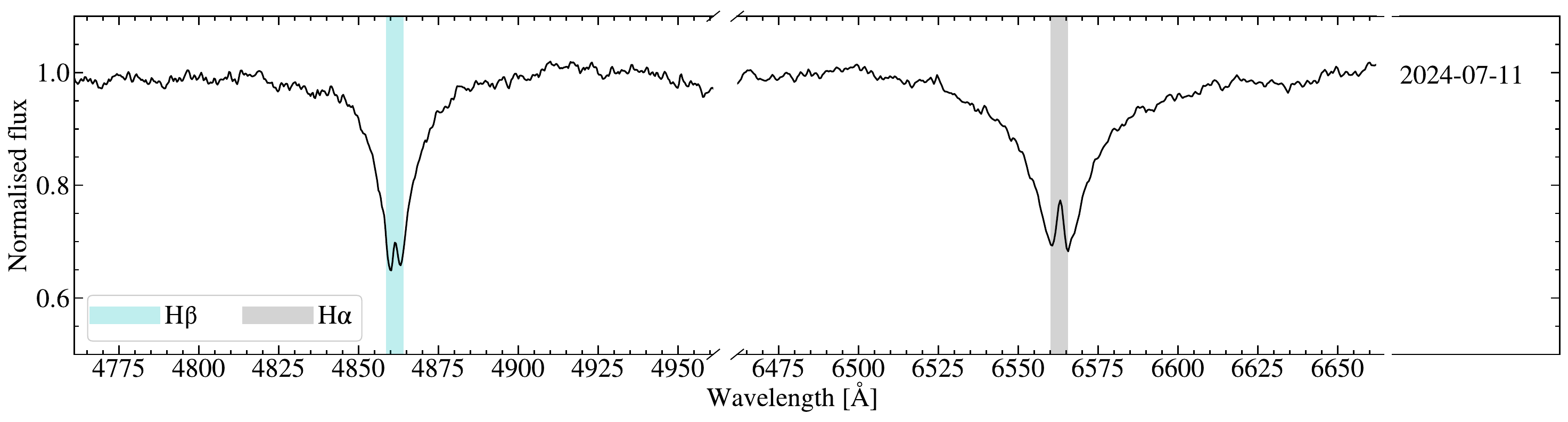}
    \caption{The stacked spectrum of WD\,J1653$-$1001, constructed from three consecutive exposures taken with the \textbf{MagE} spectrograph on 2024-07-11, around the H$\alpha$ and H$\beta$ Balmer line regions. The observation UT date is shown on the right of the plot. The spectrum is convolved with a Gaussian with a FWHM of 1\,\AA.}
    \label{fig:MagE_spectra}
\end{figure*}

We constructed a stacked spectrum of WD\,J1653$-$1001 from the three consecutive exposures taken with the MagE instrument (Section~\ref{sec:MagE}) and present it in Figure~\ref{fig:MagE_spectra}. Coverage of the H$\alpha$ and H$\beta$ Balmer line regions were achieved and clear emission in the Balmer line cores is visible.

\begin{figure}
	\includegraphics[width=\columnwidth]{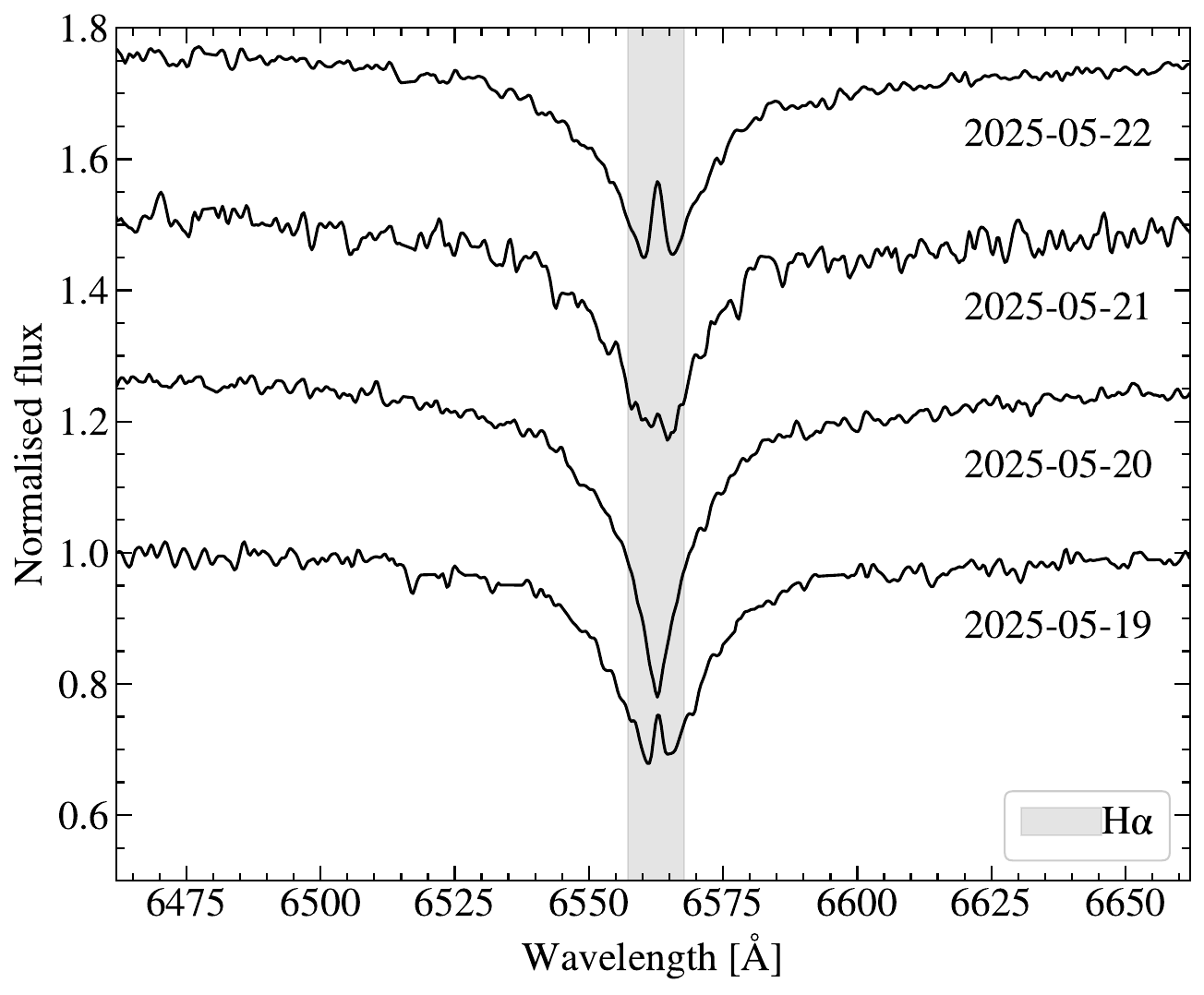}
    \caption{Four spectra of WD\,J1653$-$1001 taken with the \textbf{Binospec} spectrograph on consecutive nights from 2025-05-19 to 2025-05-22 around the H$\alpha$ Balmer line region. The observation UT date is shown on the right of the plot. All spectra are convolved with a Gaussian with a FWHM of 1\,\AA.}
    \label{fig:Binospec_spectra}
\end{figure}

Figure~\ref{fig:Binospec_spectra} shows four spectra of WD\,J1653$-$1001 taken with the Binospec spectrograph on the MMT (Section~\ref{sec:Binospec}). Coverage of the H$\alpha$ Balmer line region was achieved and clear emission strength variability is evident in the Balmer line cores over the four consecutive nights.

\begin{figure*}
	\includegraphics[width=2\columnwidth]{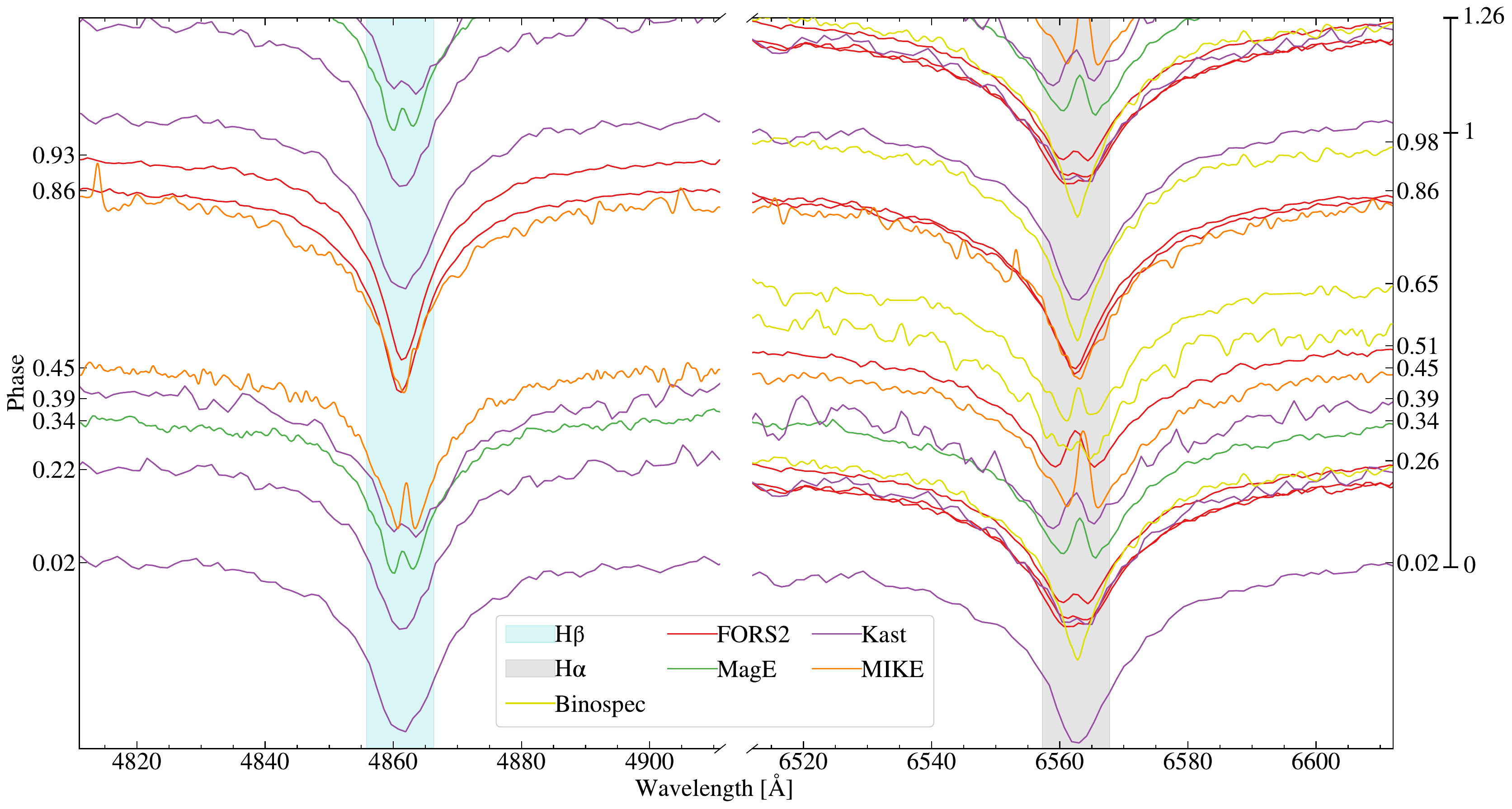}
    \caption{Phase-resolved spectra of WD\,J1653$-$1001 obtained with the Binospec, FORS2, MagE, Kast and MIKE instruments zoomed-in on the H$\alpha$ and H$\beta$ Balmer line regions. Spectra are phased to the best-fitting period from the simultaneous $g$-, $r$-, $c$- and $o$-band MCMC ($P = 80.3070$\,h). Notable phases are labeled between $\phi = 0-1$, with early-phase spectra repeated to emphasise the cyclic nature of the emission. All spectra are convolved with a Gaussian with a FWHM of 1\,\AA.}
    \label{fig:Wavelength_vs_phase}
\end{figure*}

The rotation phase of each spectroscopic observation was computed using the ephemeris in Eq.~\ref{ZTF_ATLAS_photometric_ephemeris} as $T_0$ and the best-fitting period from the simultaneous $g$-, $r$-, $c$- and $o$-band MCMC ($P = 80.3070$\,h). The H$\alpha$ and H$\beta$ emission cores from the Binospec, FORS2, MagE, Kast and MIKE spectra are shown in Figure~\ref{fig:Wavelength_vs_phase} as a function of phase, to visualise the phase-dependent variability in the line profiles. The spectra from the IDS instrument are not comparable to the other instrument spectra due to the IDS being of lower S/N, therefore they are not shown on Figure~\ref{fig:Wavelength_vs_phase}.

\begin{figure}
	\includegraphics[width=\columnwidth]{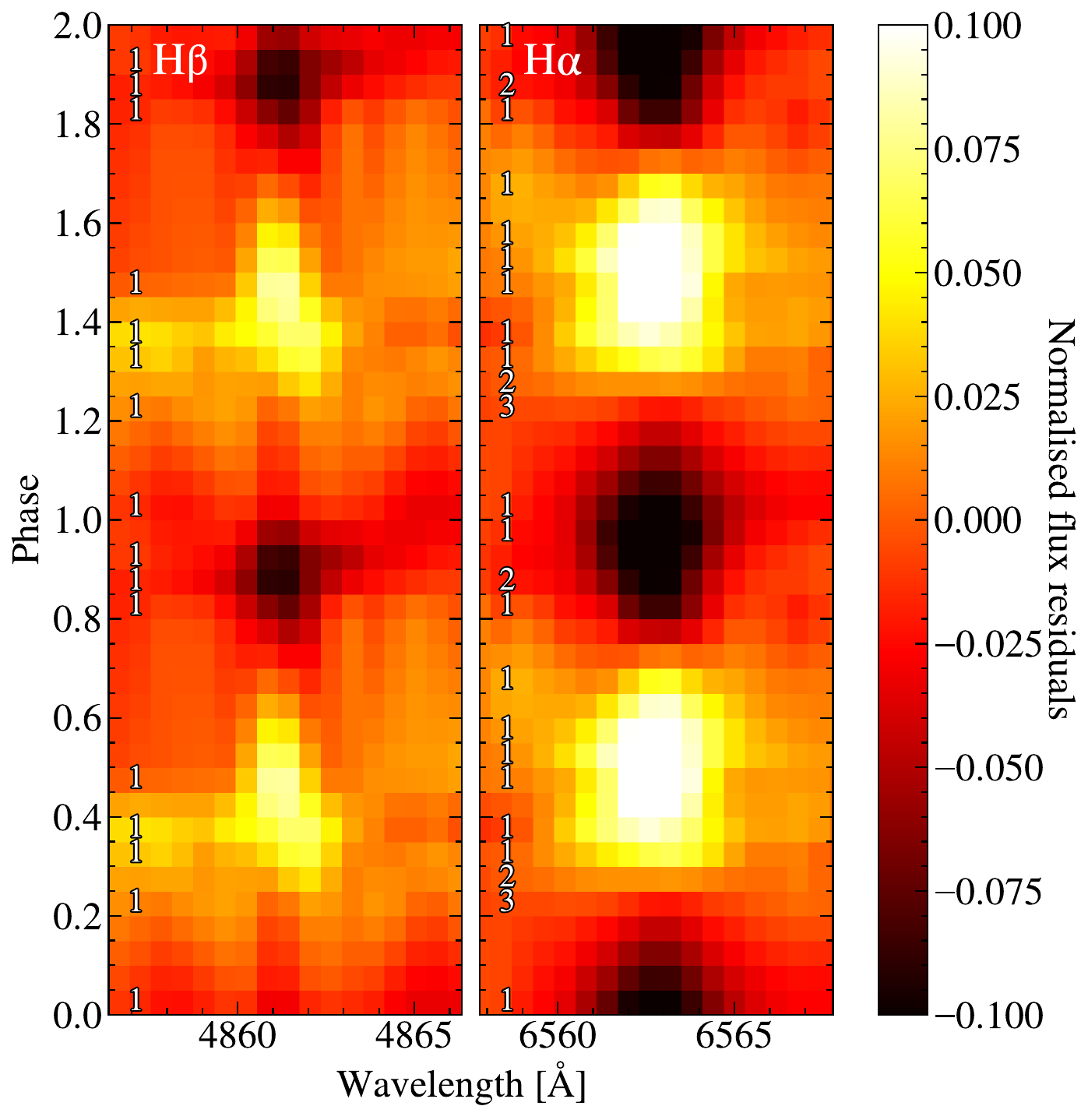}
    \caption{Trailed spectrogram of WD\,J1653$-$1001, created from the spectra obtained with the Binospec, FORS2, MagE, Kast and MIKE instruments, zoomed-in on the H$\alpha$ and H$\beta$ Balmer line regions. Spectra are phased to the best-fitting period from the simultaneous $g$-, $r$-, $c$- and $o$-band MCMC ($P = 80.3070$\,h). Lighter colours (yellow–white) indicate stronger emission, while darker colours (red–black) indicate weaker emission. The amount of spectra per phase bin are written on the left-hand side of each panel. The data are repeated over two phases for illustrative purposes.}
    \label{fig:Trailed_spectrogram}
\end{figure}

We created a trailed spectrogram using the spectra in Figure~\ref{fig:Wavelength_vs_phase} to clearly show the changes in flux around the H$\alpha$ and H$\beta$ cores, which is shown in Figure~\ref{fig:Trailed_spectrogram}. The spectra are interpolated onto the 2023-05-28 MIKE spectral wavelength grid and grouped into 20 equally spaced phase bins covering one full cycle ($\phi = 0-1$). Within each bin, we computed the average normalised spectrum to reduce noise and highlight coherent trends. As we do not have complete spectroscopic phase coverage, we fill in missing bins by interpolating the fluxes along the phase axis using the \texttt{scipy.interpolate.interp1d} function in \texttt{python}. This approach provides a continuous representation of the spectral variability with phase for WD\,J1653$-$1001, while preserving the overall flux variations observed in the data.

\begin{figure}
	\includegraphics[width=\columnwidth]{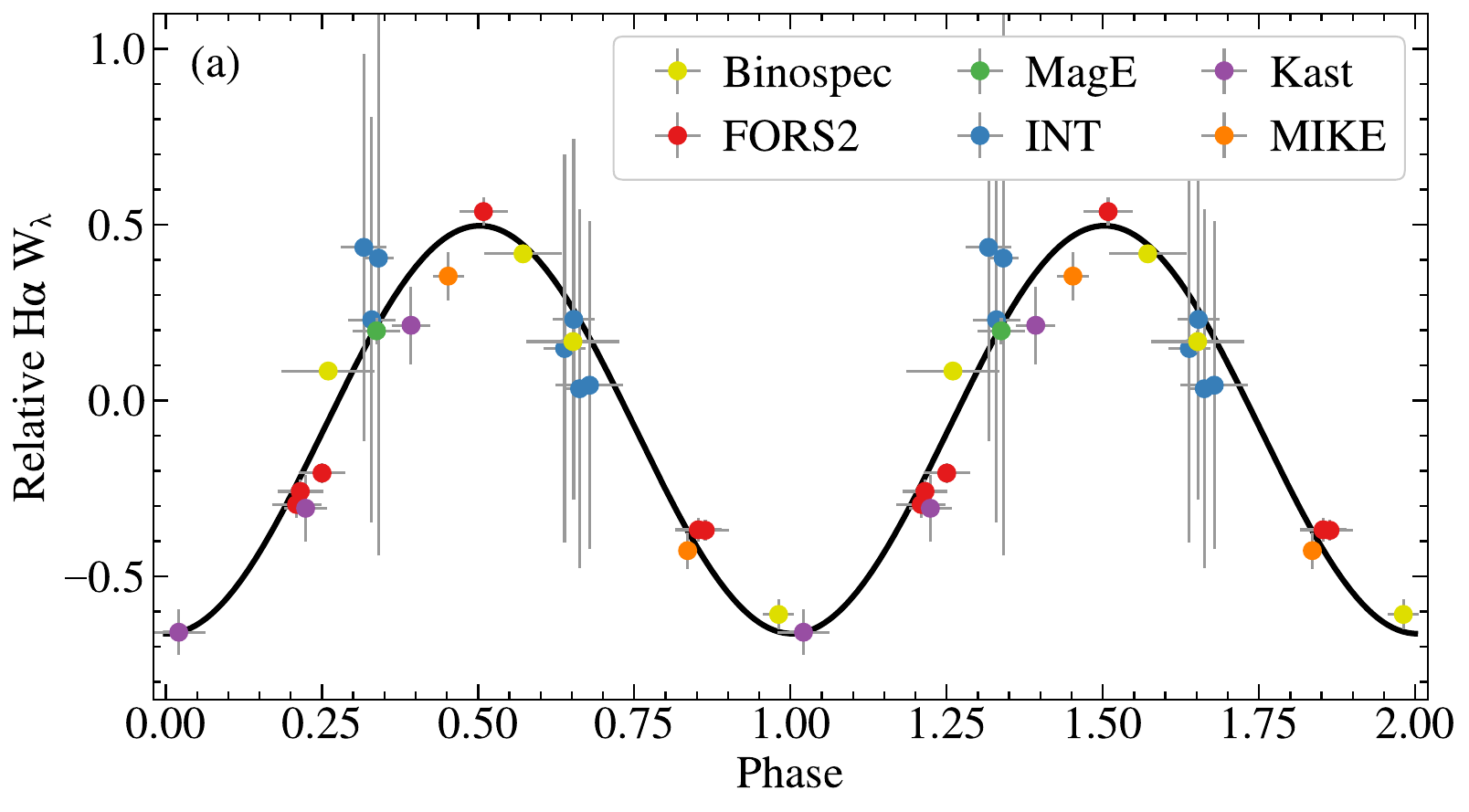}
	\includegraphics[width=\columnwidth]{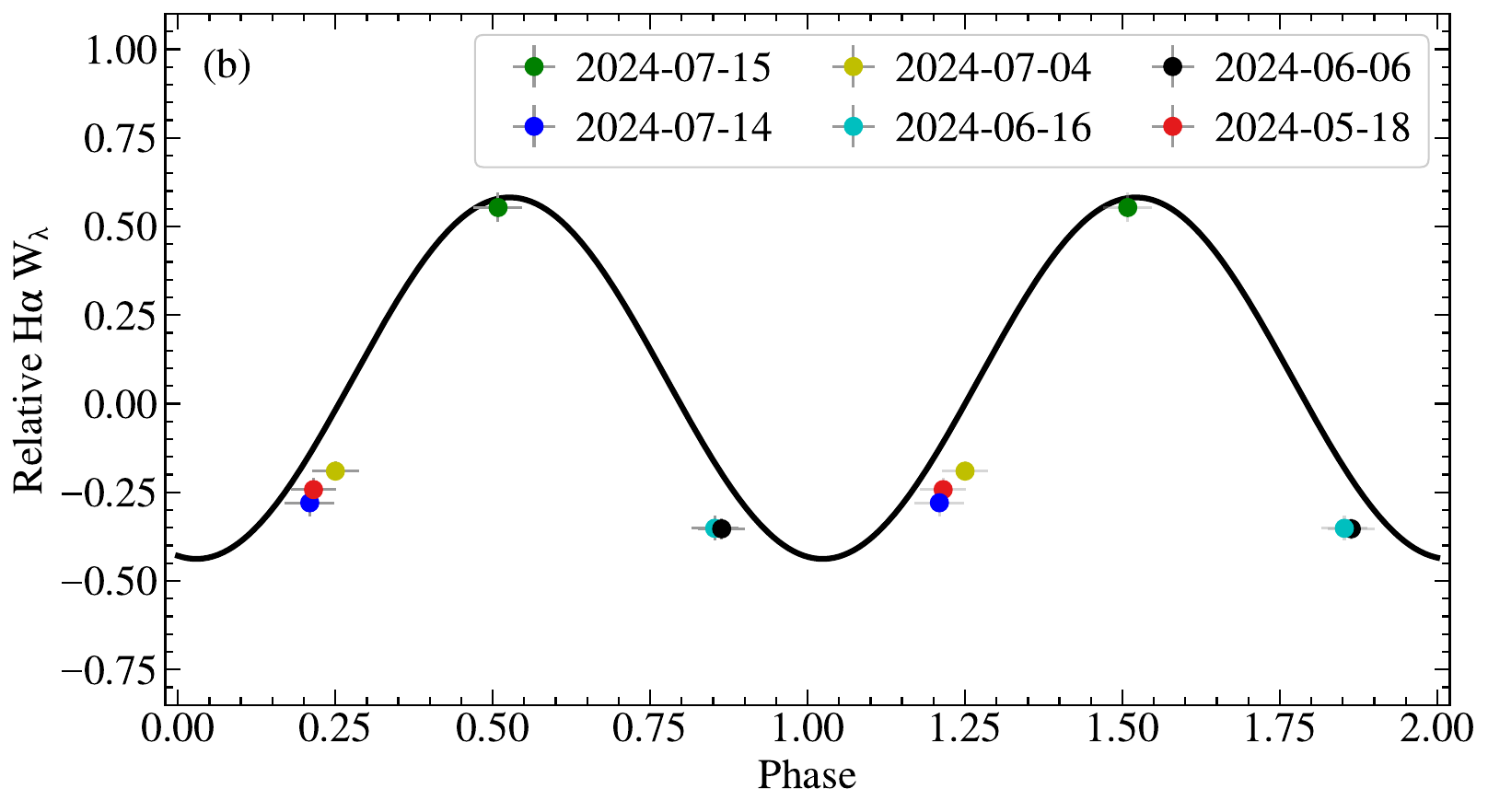}
    \caption{The phase of WD\,J1653$-$1001 as a function of the relative equivalent width (\EW) of its spectroscopically observed H$\alpha$ Balmer lines by the (a) Binospec (yellow), FORS2 (2024 data; red), MagE (green), INT (blue), Kast (purple) and MIKE (orange) telescopes/instruments and (b) six different nights of FORS2 spectra. The data are fitted with a sine wave (black overlay) and repeated over two phases for illustrative purposes. Strongest emission corresponds to the largest \EW. Error bars correspond to 3$\sigma$ uncertainties.}
    \label{fig:phase_EW_Ha}
\end{figure}

\begin{figure}
	\includegraphics[width=\columnwidth]{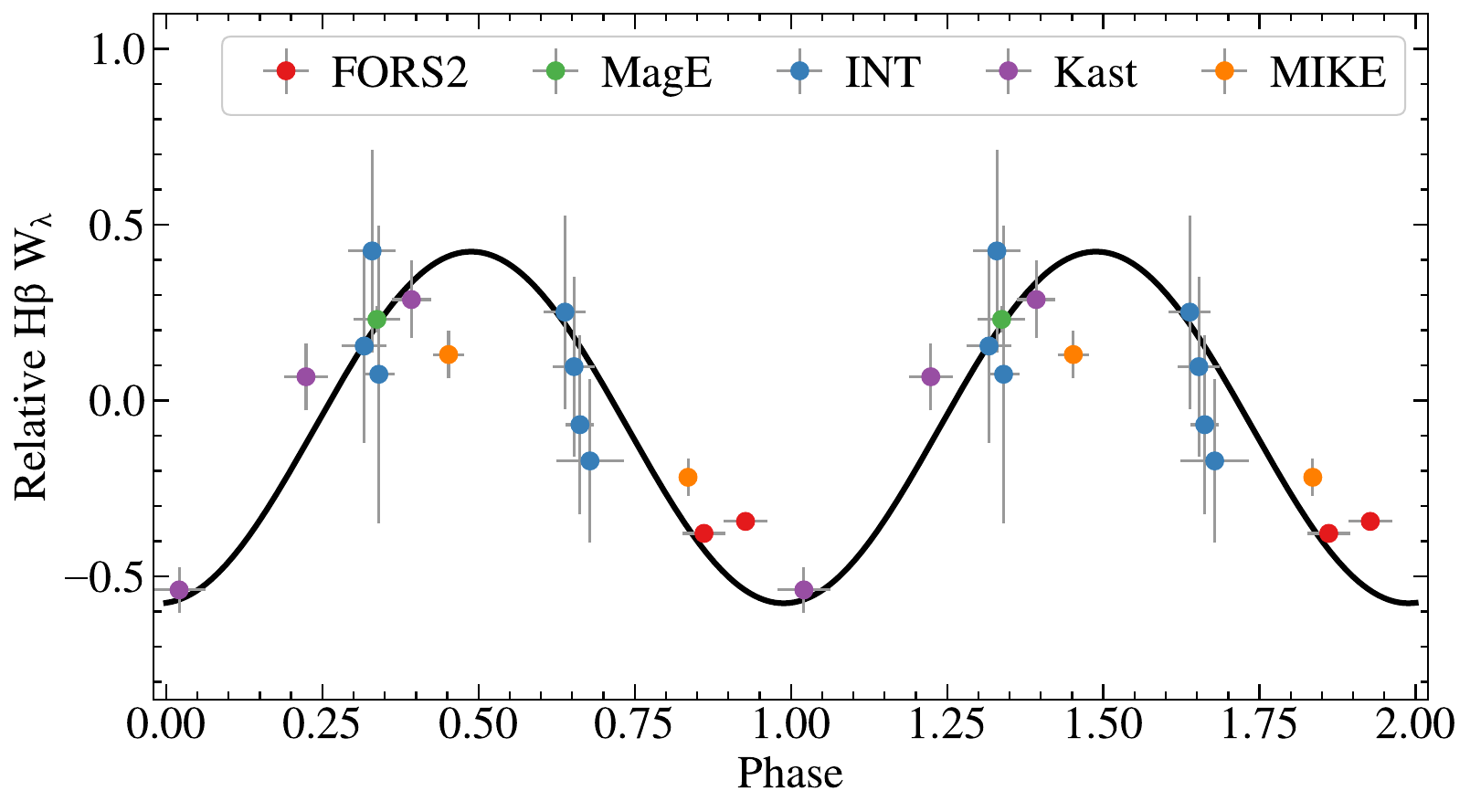}
    \caption{The phase of WD\,J1653$-$1001 as a function of the relative equivalent width (\EW) of its spectroscopically observed H$\beta$ Balmer lines by the FORS2 (2022 data; red), MagE (green), INT (blue), Kast (purple) and MIKE (orange) telescopes/instruments. The data are fitted with a sine wave (black overlay) and repeated over two phases for illustrative purposes. Strongest emission corresponds to the largest \EW. Error bars correspond to 3$\sigma$ uncertainties.}
    \label{fig:phase_EW_Hb}
\end{figure}

To investigate the emission activity of WD\,J1653$-$1001, we measured the equivalent width (\EW) of the H$\alpha$ and H$\beta$ lines in all exposures using a custom \texttt{python} routine built on the \texttt{specutils.analysis} package. The method determines the integration region adaptively: starting from an initial half-width of 2.5\,\AA\ around the line center (6562.79\,\AA\ for H$\alpha$ and 4861.35\,\AA\ for H$\beta$), the window is iteratively expanded in 0.5\,\AA\ steps until the flux at its edges approaches the continuum level (within 2 per cent); or, until a maximum half-width of 4.0\,\AA\ is reached. This ensures the \EW\ calculation captures the full emission core while minimising contamination from noise in the line wings. The continuum level was taken to be 1.0, consistent with our normalised spectra. Uncertainties on \EW\ were estimated via $10^3$ Monte Carlo iterations that perturbed the flux according to the propagated flux errors.

Since the emission strength varies with phase, we found the difference in \EW\ between the observed line cores and the line core which has the weakest emission. This method ensures larger \EW\ values directly correspond to stronger emission. The values were then normalised by subtracting the mean, yielding our final \EW\ values. Figure~\ref{fig:phase_EW_Ha}(a) shows how the \EW\ of the H$\alpha$ line cores vary with phase in the Binospec, 2024 FORS2, MagE, stacked INT, Kast and MIKE observed spectra, and Figure~\ref{fig:phase_EW_Ha}(b) shows the six individual 2024 FORS2 \EW\ measurements. Figure~\ref{fig:phase_EW_Hb} shows how the \EW\ of the H$\beta$ line cores vary with phase in the 2022 FORS2, MagE, stacked INT, Kast and MIKE observed spectra.

The fitted sine waves in Figures~\ref{fig:phase_EW_Ha}~and~\ref{fig:phase_EW_Hb} include period as a free parameter. The resulting best-fitting periods from the H$\alpha$ \EW\ measurements is $P = 80.53 \pm 0.34$\,h, and from the H$\beta$ \EW\ measurements is $P = 78.50 \pm 0.60$\,h. The H$\alpha$ period is within $1\sigma$ of the photometric period from the simultaneous $g$-, $r$-, $c$- and $o$-band MCMC ($P = 80.3070$\,h), and the H$\beta$ period is within $3\sigma$. The larger discrepancy between the photometric period and the H$\beta$ period is likely due to the smaller number of data points and the larger uncertainties compared to H$\alpha$.  

Furthermore, we searched for periodic modulation in the H$\alpha$ \EW\ using a Lomb-Scargle periodogram. The periodogram was computed from the BMJD–50\,000 times of the spectra and their corresponding \EW\ measurements, with uncertainties included as weights. Similar to our ZTF and ATLAS time series analysis, we only searched for the strongest unique signals at periods $< 28$\,d, separated by $\geq 1$\,h and above a FAP threshold of $5\sigma$. Only one signal satisfied these conditions, at a period of $P = 80.2922$\,h. To quantify the uncertainty on the spectroscopic period, we performed a Monte Carlo perturbation of the \EW\ data by their measurement uncertainties. With 2000 iterations, this approach gives a consistent value of $P = 80.2922 \pm 0.0108$\,h. This spectroscopic period is within $2\sigma$ of the H$\alpha$ sine-fit result from Figure~\ref{fig:phase_EW_Ha} and the photometric period from the simultaneous $g$-, $r$-, $c$- and $o$-band MCMC. 

\subsection{Spectropolarimetric analysis}
\label{sec:Spectropolametric analysis}

\begin{figure*}
	\includegraphics[width=2\columnwidth]{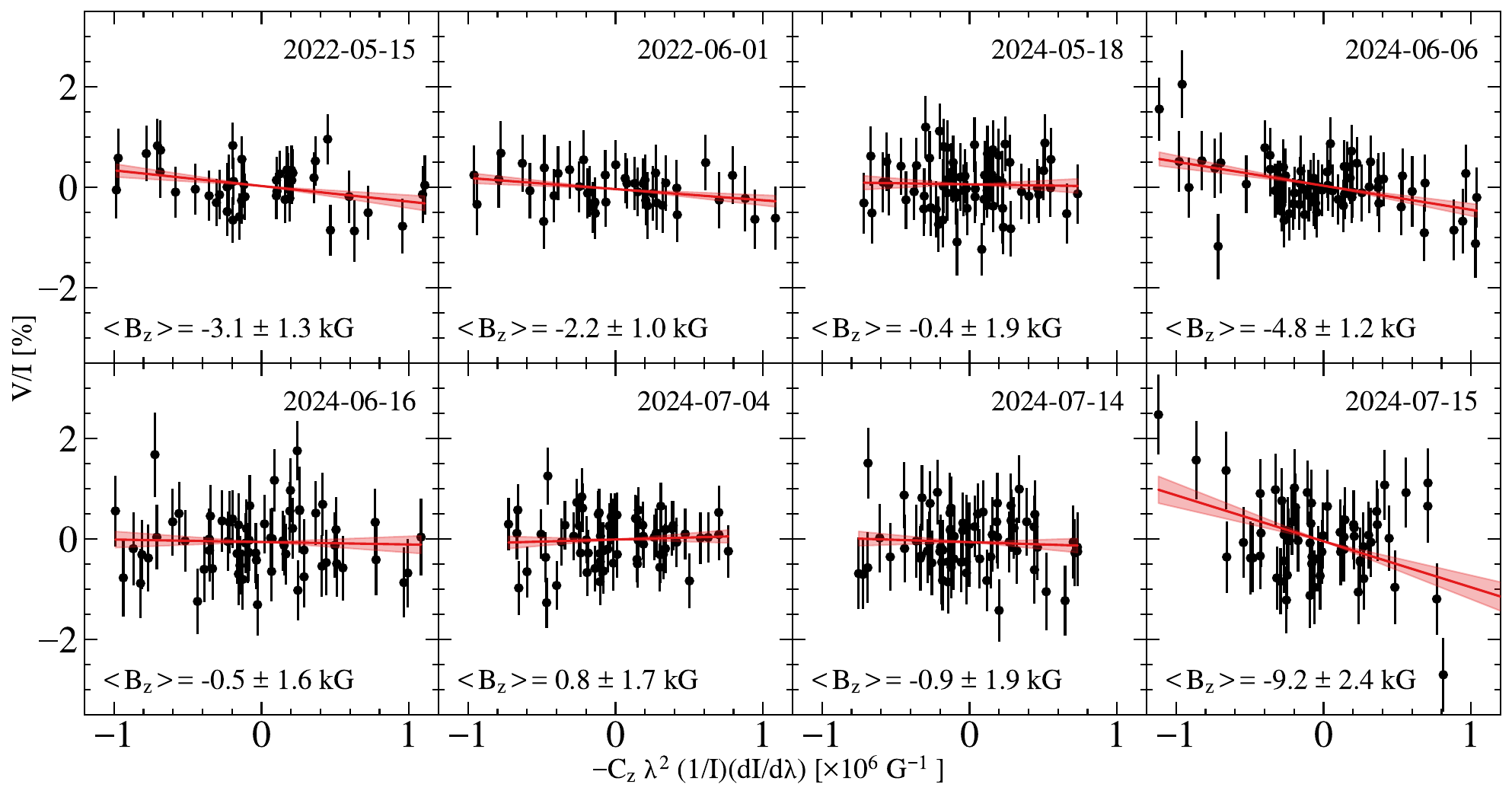}
    \caption{The best-fitting linear regression to the $V/I$ profiles of the FORS2 2022 and 2024 polarimetric data using Eq.~\ref{V_over_I}. The gradient and uncertainty of the best-fitting line (red) yields $\langle B_z \rangle$, which is displayed in the lower left of each panel. The observation UT date is in the upper right of each panel. Statistically significant magnetic field detections occurred on 2024-06-06 and 2024-07-15. Marginal magnetic field detections occurred on 2022-05-15 and 2022-06-01.}
    \label{fig:Bz}
\end{figure*}

\begin{figure*}
	\includegraphics[width=2\columnwidth]{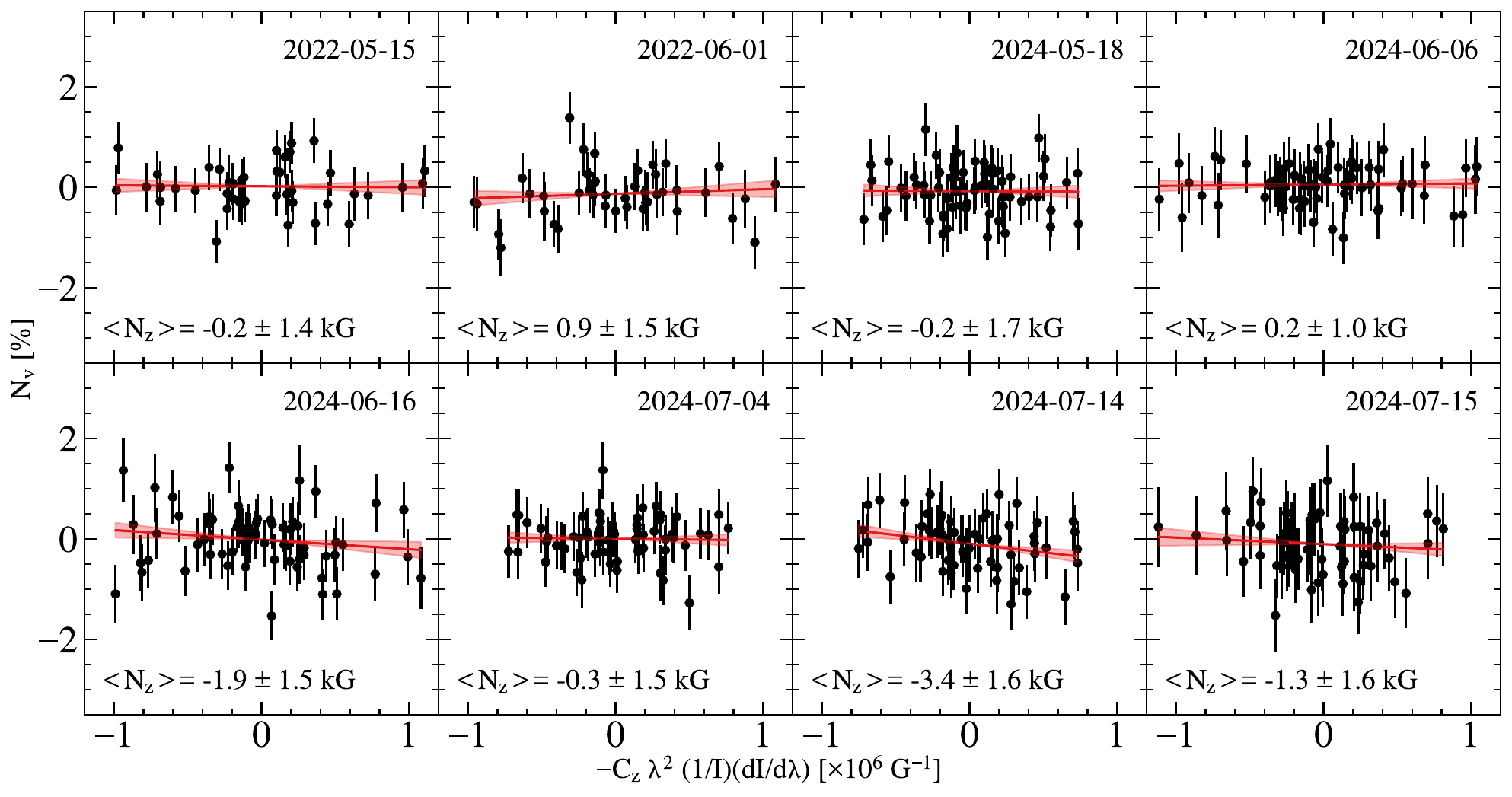}
    \caption{Same as Figure~\ref{fig:Bz} but for the $N_V$ profiles. The gradient and uncertainty of the best-fitting line (red) yields $\langle N_z \rangle$, which is used as quality control for the corresponding $\langle B_z \rangle$ measurements. A statistically significant zero $\langle N_z \rangle$ constitutes a reliable observation. Only 2024-06-16 and 2024-07-14 have non-zero $\langle N_z \rangle$ values which indicates they comprise of slightly noisier data than the other epochs and should be treated with caution.}
    \label{fig:Nz}
\end{figure*}

\begin{table*}
    \centering
    \caption{The phase ($\phi$) and mean longitudinal field ($\langle B_z \rangle$) measurements, with $1\sigma$ uncertainty, of WD\,J1653$-$1001 from the FORS2 2022 and 2024 observations. The significance ($\sigma$) of the detection is provided, along with the p-value from a two-sided significance test and category of detection.}
    \label{tab:Bz_results}
    \begin{tabular}{cccccc}
    \hline
    \hline
    Date & Phase & $\langle B_z \rangle$ & $\sigma$ & p-value & Detection? \\
    yyyy-mm-dd & & [kG] &  &  &  \\
    \hline
    \hline
    2022-05-15 & 0.86 & $-3.11 \pm 1.31$ & 2.37 & 0.0223 & Marginal \\
    2022-06-01 & 0.93 & $-2.22 \pm 1.04$ & 2.13 & 0.0386 & Marginal \\
    \hline
    2024-05-18 & 0.22 & $-0.45 \pm 1.94$ & 0.23 & 0.819 & No \\
    2024-06-06 & 0.86 & $-4.78 \pm 1.23$ & 3.88 & $2.37\times 10^{-4}$ & Yes \\
    2024-06-16 & 0.85 & $-0.49 \pm 1.63$ & 0.30 & 0.765 & No \\
    2024-07-04 & 0.25 & $+0.81 \pm 1.69$ & 0.48 & 0.634 & No \\
    2024-07-14 & 0.21 & $-0.92 \pm 1.92$ & 0.48 & 0.633 & No \\
    2024-07-15 & 0.51 & $-9.16 \pm 2.39$ & 3.83 & $2.91\times 10^{-4}$ & Yes \\
    \hline
\end{tabular}
\end{table*}

We analysed the eight epochs of spectropolarimetric data obtained from the FORS2 instrument in 2022 and 2024 (Section~\ref{sec:FORS2}) to determine whether WD\,J1653$-$1001 has a weak magnetic field. As spectropolarimetry measures the circular polarisation of spectral lines due to a magnetic field along the line-of-sight, it yields $\langle B_z \rangle$ as the gradient from $V/I$ \citep{Bagnulo2002}. This is obtained by using the weak-field approximation, 
\begin{equation}
 \frac{V}{I} = -g_{\textnormal{eff}} C_z \lambda^2 \frac{1}{I} \frac{\textnormal{d}I}{\textnormal{d}\lambda} \langle B_z \rangle,
 \label{V_over_I}
\end{equation}
{\noindent}with
\begin{equation}
 C_z = \frac{e}{4\pi m_\textnormal{e} c^2} \simeq 4.67 \times 10^{-13}~ \angstrom^{-1} ~\textnormal{G}^{-1},
 \label{c_z}
\end{equation}
{\noindent}where $g_{\textnormal{eff}}$ is the effective Land\'e factor (which is 1.0 for Hydrogen Balmer lines; \citealt{Casini1994, Bagnulo2002}), $e$ is the electron charge, $m_{\textnormal{e}}$ is the electron mass and $c$ is the speed of light. For the Balmer lines, the weak-field approximation remains valid only for field strengths $\lesssim 10$\,kG, where the Zeeman splitting is negligible compared to the local line width. Eq.~\ref{V_over_I} was applied to each pixel in the $V/I$ profile over the wavelength range $6540 < \lambda < 6590$\,\AA\ for the H$\alpha$ region and $4846 < \lambda < 4876$\,\AA\ for the H$\beta$ region. Then, a correlation diagram of $V/I$ as a function of the resulting $-C_z \lambda^2 (1/I) (dI/d\lambda)$ values was created and a straight line fitted. The measured gradient of the straight line is $\langle B_z \rangle$. A non-zero gradient constitutes a magnetic field detection, whereas a gradient of zero means there is no magnetic field detection. This method of measuring $\langle B_z \rangle$ is common in spectropolarimetric analysis \citep[e.g.][]{Bagnulo2002, Hubrig2006, Hubrig2009b, Hubrig2009a, Bagnulo2012, Landstreet2014, Bagnulo2024a} thus is used in this paper. 

Figure~\ref{fig:Bz} presents the best-fitting linear regression to $V/I$ obtained with FORS2 in 2022 and 2024. The $\langle B_z \rangle$ measurements from each fit are shown in Table~\ref{tab:Bz_results}, along with $1\sigma$ uncertainties and the observation phase. Also shown are the significance ($\sigma$) of the $\langle B_z \rangle$ detections and the results from a two-sided significance test on the linear regression, where the $\sigma$ and p-value determines whether the measured $\langle B_z \rangle$ value constitutes a detection. 

For quality control, we applied the same regression analysis to the diagnostic $N_V$ profiles as to the $V/I$ profiles. The $N_V$ profiles are constructed by combining the same series of exposures used to form the $V/I$ profiles, but in a particular way such that any real stellar polarisation signal cancels out, leaving only noise and potential instrumental artifacts \citep{Donati1997, Bagnulo2012}. Therefore, if the data are clean and do not suffer from systematics, $N_V$ should scatter randomly around zero and $\langle N_z \rangle$ should be statistically consistent with zero. As such, $N_V$ serves as a crucial diagnostic for assessing the reliability of $\langle B_z \rangle$ detections. The uncertainties assigned to $N_V$ were taken to be identical to those of $V/I$, since the two quantities have equal variance at each pixel. Following \citet{Bagnulo2012}, we then rescaled the uncertainties such that the distribution of $N_V/\sigma_{N_V}$ has approximately unit variance. The gradient from the linear regression fit yields the $\langle N_z \rangle$ values. Figure~\ref{fig:Nz} shows the $N_V$ profiles, best-fitting regression lines and $\langle N_z \rangle$ values for the 2022 and 2024 FORS2 observations. 

We also estimate the magnetic field strength of WD\,J1653$-$1001 using our eight epochs of $\langle B_z \rangle$ measurements, regardless of detection category. For simplicity, we assume a dipolar magnetic field configuration with a polar strength, $B_d$. We compute the root-mean-square of $\langle B_z \rangle$,
\begin{equation}
 \langle B_z \rangle_{\mathrm{rms}} = \left(\frac{1}{N} \sum_{i}^{N} \langle B_z \rangle ^2 _i \right)^{1/2},
 \label{Bz_rms}
\end{equation}
{\noindent}which gives $3.93$\,kG. According to Table~1 of \citet{Kochukhov2024}, this value would likely correspond to a median $B_d$ of $22$\,kG. Adopting a $\pm 2\sigma$ confidence interval (95.5 per cent), we estimate that $B_d$ lies between $15$\,kG and $54$\,kG. Table~\ref{tab:stellar parameters} presents the median $B_d$ and $2\sigma$ errors. 

\section{Discussion}
\label{sec:Discussion}

\subsection{Validation of the \texorpdfstring{80\,h}{80 h} period}
\label{sec:validation of period}
All analyses in this work consistently identify a period of around $80.3$\,h for WD\,J1653$-$1001 across independent datasets and methods (Table~\ref{tab:measured_periods}). For the ZTF photometry, the MCMC analysis derived periods of $80.3006 \pm 0.0041$\,h in the $g$-band, $80.3000 \pm 0.0042$\,h in the $r$-band, and $80.2994 \pm 0.0042$\,h in the the combined band. These are all in excellent agreement of within $1\sigma$. Similarly, the periods obtained from least-squares sine fits on the ZTF individual and combined band light curves are also all within $1\sigma$. The corresponding MCMC and sine fit periods for the $g$-, $r$- and combined bands are all within $1\sigma$ too, indicating that the $80.3$\,h modulation is robustly present in the ZTF data. 

The ATLAS photometry independently reproduces the periodicity measured with ZTF and at a higher precision. The period values obtained from our MCMC analysis of the $c$-band is $80.3066 \pm 0.0029$\,h, $o$-band is $80.3080 \pm 0.0028$\,h and the combined band is $80.3077 \pm 0.0021$\,h. These are all mutually consistent to well below $1\sigma$. Similarly, the periods obtained from a least squares fit to a sinusoid on the ATLAS individual and combined band light curves are all within $1\sigma$. The corresponding MCMC and fit sine curve periods for the $c$-, $o$- and combined bands are all within $1\sigma$ too.

Performing a simultaneous MCMC fit of all four datasets ($g$, $r$,  $c$ and $o$) yields a period of $80.3070 \pm 0.0007$\,h and a sinusoidal fit yields $80.3073 \pm 0.0007$\,h. These two periods from independent techniques are statistically indistinguishable, differing by $0.3\sigma$. These are highly precise measurements of the period so their agreement strongly validates the detected period. 

The additional photometric data from the Calar Alto telescope and ASAS-SN (Section~\ref{sec:Additional photometry}) corroborate the period detected from ZTF and ATLAS. The best-fitting periods from both the MCMC and sinusoidal fits agree within $1\sigma$ (Section~\ref{sec:Photometric variability of WDJ1653}), and are consistent with the best-fitting period from the simultaneous $g$-, $r$-, $c$- and $o$-band MCMC to the same degree.

The spectroscopic period derived from the H$\alpha$ \EW\ ($P = 80.2922 \pm 0.0108$\,h; Section~\ref{sec:Spectroscopic variability}) is within $2\sigma$ of the photometric period obtained from the simultaneous $g$-, $r$-, $c$- and $o$-band MCMC (Section~\ref{sec:Photometric variability of WDJ1653}). This concordance between photometric and spectroscopic periods further validates $P \approx 80.3$\,h as the true orbital period of WD\,J1653$-$1001, rather than an alias or spurious signal.

These results confirm and refine the tentative period reported in \citet{Elms2023} from ZTF photometry, which at the time was most likely $P \approx 80.31$\,h. The authors used earlier ZTF DR15 data, of which analysis was limited by contamination from the background main-sequence star \textit{Gaia} DR3 4334641562479650816. Over the 4\,years 8\,months baseline of ZTF DR15, WD\,J1653$-$1001 and the contaminant star had angular separations of 0.971" to 0.832". ZTF DR23 provides an extended baseline of an additional 1\,year 11\,months, which increased the angular separation between WD\,J1653$-$1001 and the contaminant star to 1.214". Thus, the ZTF data for WD\,J1653$-$1001 has become less contaminated since it was last analysed in \citet{Elms2023}. As a result, the data has improved in quality over the past couple of years but retained the same strongest signal. The agreement between ZTF and ATLAS - two independent surveys with different instruments, passbands, and cadences - provides robust evidence that the $80.3$\,h modulation is intrinsic to WD\,J1653$-$1001.

\subsection{Antiphase photometric and spectroscopic variability}
\label{sec:antiphase_relationship}
To phase the photometric and spectroscopic data, we chose epochs of $T_\mathrm{0}$ such that $\phi = 0$ corresponds to the photometric maximum in the light curves of WD\,J1653$-$1001 (Section~\ref{sec:Photometric variability of WDJ1653} and Figure~\ref{fig:ZTF_ATLAS_lc_phase}). The phased spectra from Binospec, FORS2, MagE, Kast, and MIKE, shown in Figure~\ref{fig:Wavelength_vs_phase}, reveal pronounced phase-dependent variability in the H$\alpha$ and H$\beta$ Balmer line cores. The emission strength clearly increases and decreases throughout the phase, with the strongest emission observed around $\phi \approx 0.35-0.55$. At phases nearing photometric maximum, $\phi \approx 0.8-1.0$, the emission weakens, indicating a clear antiphase relationship between the photometric flux and emission strength. 

We further visualised this behaviour using the trailed spectrogram shown in Figure~\ref{fig:Trailed_spectrogram}, which presents the spectra flux residuals as a function of phase and wavelength. To account for incomplete spectroscopic phase coverage, the fluxes were interpolated along the phase axis, providing a continuous view of the spectral variability across the period. This highlights the distinct strong and weak emission phases observed in WD\,J1653$-$1001, and demonstrates that the observed modulation is intrinsic rather than an artefact of our incomplete spectroscopic sampling. 

The spectroscopic emission modulation is also presented in Figure~\ref{fig:phase_EW_Ha}, this time with the H$\alpha$ \EW\ as a function of phase for the Binospec, 2024 FORS2, MagE, stacked INT, Kast and MIKE spectra. This reveals weakest emission (smallest \EW) corresponds to $\phi \approx 0$. The same relationship is evident for the H$\beta$ \EW\ as a function of phase in Figure~\ref{fig:phase_EW_Hb}.  

The spectroscopic variability of WD\,J1653$-$1001 throughout its phase is clearly demonstrated in the 2024 FORS2 data in Figures~\ref{fig:FORS2_2024_spectra}~and~\ref{fig:phase_EW_Ha}(b). The variation of the H$\alpha$ emission core is visually evident in the six epochs observed with FORS2 in 2024 in Figure~\ref{fig:FORS2_2024_spectra}. These same six epochs are distinguished in Figure~\ref{fig:phase_EW_Ha}(b), which presents the measured \EW\ of the H$\alpha$ emission core as a function of phase. Together, these figures illustrate the phase-dependent evolution of the H$\alpha$ emission core, with weakest emission occurring at $\phi \approx 0$, captured consistently with a single instrument.

To quantify the antiphase relationship between photometric flux and emission, we repeated the sinusoidal fit in Figure~\ref{fig:phase_EW_Ha}(a), but this time fixing the period to the photometric period $80.3073 \pm 0.0007$\,h. Converting the best-fitting phase to fractional cycles gives the maximum H$\alpha$ \EW\ at $\phi = 0.503 \pm 0.002$ and a minimum at $\phi = 0.003 \pm 0.002$, thus the H$\alpha$ \EW\ minimum almost perfectly coincides with the photometric maximum ($\phi = 0.0$), within uncertainties.

The evidence that the photometric and spectroscopic variability of WD\,J1653$-$1001 occurs at the same period, but in antiphase, strongly suggests a single physical mechanism is causing the variability. An explanation that supports this evidence is that the star hosts a photospheric dark spot/region with a temperature-inverted and optically thin chromospheric emission region \citep{Walters2021}. In this scenario, the observed variability arises naturally from stellar rotation, with flux deficits and Balmer line emission occurring periodically by the changing visibility of the photospheric spot/region.

One curious feature that emerges from Figure~\ref{fig:Trailed_spectrogram} is that the H$\alpha$ emission appears to show a slight phase offset relative to H$\beta$. Similar to above, repeating the sinusoidal fit in Figure~\ref{fig:phase_EW_Hb}, but fixing the period to the photometric period $80.3073 \pm 0.0007$\,h, reveals the maximum H$\beta$ \EW\ occurs at $\phi = 0.498 \pm 0.005$ and a minimum at $\phi = 0.998 \pm 0.005$. This implies an apparent phase offset between H$\alpha$ and H$\beta$ of $\phi \approx 0.005$ -- however, the visual offset in Figure~\ref{fig:Trailed_spectrogram} seems $\approx 10$ times larger than this. 

If real, this offset could provide insight into the geometry and structure of the proposed photospheric spot/region.
Because H$\alpha$ and H$\beta$ form at different depths within the stellar atmosphere/chromosphere, they effectively probe distinct vertical regions of the emitting layer. At those depths, the properties of the emitting spot/region may vary in terms of its wavelength dependence or horizontal physical size, leading to subtle differences when individual Balmer line emission forms over the rotation period. 
This interpretation is further supported by Figure~\ref{fig:Wavelength_vs_phase}, which shows that the H$\alpha$ and H$\beta$ line profiles exhibit subtle structural differences at comparable phases. However, it is also possible that the apparent phase offset is not intrinsic to the star, but instead arises from observational or instrumental effects. The data used in this analysis were obtained using multiple instruments with heterogeneous spectrographs and differing instrumental setups, where small inconsistencies in wavelength calibration, imperfect calibration corrections, or the use of different gratings to observe H$\alpha$ and H$\beta$ regions, could all introduce apparent phase shifts between the lines. Additionally, the H$\beta$ data in particular required more extensive interpolation in Figure~\ref{fig:Trailed_spectrogram}, due to sparse spectroscopic sampling. These effects could collectively mimic a phase offset between the lines so we are cautious to confirm this as a real feature. To verify whether the offset is genuine, future high-resolution and phase-resolved spectroscopic observations, ideally  spanning a complete orbital cycle, are required.

\subsection{Detection of a weak magnetic field}
\label{sec:detection of a magnetic field}
Using spectropolarimetric observations obtained with FORS2 in 2022 and 2024, it is possible to measure the $\langle B_z \rangle$ of WD\,J1653$-$1001 at multiple epochs (see Section~\ref{sec:Spectropolametric analysis} for details). A statistically significant non-zero $\langle B_z \rangle$ measurement constitutes a definitive detection of a magnetic field, even though Zeeman-splitting is absent.  Figure~\ref{fig:Bz} shows that the $V/I$ profiles in the eight epochs of FORS2 data have gradients ranging from $\langle B_z \rangle~=~-9.16~\pm~2.39$\,kG to $\langle B_z \rangle~=~0.81~\pm~1.69$\,kG, with varying levels of significance. 

To help determine the significance of each $\langle B_z \rangle$ measurement, and given that the inferred magnetic field strengths are of order $1$\,kG, we report both the $\sigma$ of each $\langle B_z \rangle$ detection and p-value from a two-sided significance test in Table~\ref{tab:Bz_results}. The null hypothesis of our significance test corresponds to a zero gradient, i.e. no correlation between $V/I$ and $-C_z \lambda^2 (1/I) (dI/d\lambda)$, which implies the absence of a magnetic field. The alternative hypothesis is that the gradient is non-zero, corresponding to the presence of a magnetic field. We set strict detection criteria, classifying a measurement as a definite detection when $\sigma \geq 3$ and $p \leq 0.01$, meaning we have a $\leq 1$ per cent chance of accepting a false positive. Measurements with $2 \leq \sigma < 3$ and $0.01 \leq p < 0.05$ are classified as marginal detections. All other measurements outside these criteria are considered non-detections. This strict regime, which is more stringent than the conventional $\alpha = 0.05$ threshold, follows common practice in the analysis of stellar spectropolarimetry \citep[e.g.][]{Hubrig2009a, Hubrig2009b, Bagnulo2012, Landstreet2014}, as it reduces the probability of spurious magnetic field detections arising from noise fluctuations and small instrumental instabilities. 

Two of the 2024 epochs yield statistically significant magnetic field detections: the 2024-06-06 data has $\langle B_z \rangle = -4.78 \pm 1.23$\,kG at $3.88\sigma$, with $p = 0.000237$; the 2024-07-15 data has $\langle B_z \rangle = -9.16 \pm 2.39$\,kG at $3.83\sigma$, with $p = 0.000291$. The remaining four epochs of 2024 FORS2 data yield $\langle B_z \rangle$ measurements outside our detection criteria, thus are consistent with non-detections. 
 
The FORS2 data from 2022-05-15 has a measured $\langle B_z \rangle$ of $-3.11 \pm 1.31$\,kG at $2.37\sigma$, with $p = 0.0223$. The data from 2022-06-01 has a measured $\langle B_z \rangle$ of $-2.22 \pm 1.04$\,kG at $2.13\sigma$, with $p = 0.0386$. Both of these epochs from 2022 yield marginal magnetic field detections.

To check that the FORS2 $V/I$ data are clean, and therefore the reliability of the corresponding $\langle B_z \rangle$ measurements, we analysed the $N_V$ of the data. For clean data, $N_V$ should be consistent with zero within our H$\alpha$ and H$\beta$ wavelength regions (see Section~\ref{sec:Spectropolametric analysis}) and should not exhibit significant structure. The gradient of the linear regression of the $N_V$ profiles yields the $\langle N_z \rangle$ values, which provide a quantitative measure of any spurious signals. Figure~\ref{fig:Nz} shows the $N_V$ profiles, best-fitting regression lines and $\langle N_z \rangle$ values for the 2022 and 2024 FORS2 observations. Six out of eight FORS2 epochs have $\langle N_z \rangle$ values consistent with zero, including the four epochs which produced significant or marginal magnetic field detections. This validates the magnetic field detections from the 2022 and 2024 FORS2 data sets. The remaining two FORS2 epochs have non-zero $\langle N_z \rangle$ values: 2024-06-16 has $\langle N_z \rangle = -1.9 \pm 1.5$\,kG and 2024-07-14 has $\langle N_z \rangle = -3.4 \pm 1.6$\,kG. Therefore, these observations are slightly noisier than the other epochs and should be treated with caution, though both measurements are below 3$\sigma$ significance and the corresponding $\langle B_z \rangle$ measurements for these two epochs are non-detections. 

Out of the eight FORS2 $\langle B_z \rangle$ measurements, seven are negative. The remaining $\langle B_z \rangle$ measurement, from 2024-07-04, is positive but consistent with zero. In total, four $\langle B_z \rangle$ measurements are consistent with zero. The other four $\langle B_z \rangle$ measurements are non-zero, negative, and constitute detections: 2022-05-15 and 2022-06-01 are marginal detections; 2024-06-06 and 2024-07-15 are significant detections. This suggests that if WD\,J1653$-$1001 has a dipole field configuration with two magnetic poles, only the negative pole has been observed. The most likely explanation for this is the magnetic axis is tilted relative to the line-of-sight, and the star rotates in such a way that the opposite pole never becomes visible. Therefore, the ZTF and ATLAS measurements of the period of variability in Table~\ref{tab:measured_periods} are likely the rotation period. This is further evidence against the hypothesis that the true period of variability could be $2P$ (Section~\ref{sec:Photometric variability of WDJ1653}). 

Furthermore, we used the 2024 spectropolarimetric data to check for high frequency variations in the epochs, in case the true rotation period is shorter than those listed in Table~\ref{tab:measured_periods}. Each FORS2 data set consists of eight subexposures spanning $1$\,h in total, which are recombined to obtain unique and robust Stokes $I$ and Stokes $V/I$ profiles. As the period of variability from ZTF and ATLAS measurements is $\approx 80.3$\,h, no significant variation in the H$\alpha$ region is expected within individual epochs. We analysed the Stokes $I$ spectra of the individual subexposures, spaced $\approx 6$\,min apart, from the 2024-06-06 (phase of weak emission) and 2024-07-15 (phase of strong emission) epochs. No discernible variation in the H$\alpha$ region was evident, which reinforces the conclusion that WD\,J1653$-$1001 has a rotation period significantly longer than $1$\,h. 

In Section~\ref{sec:Spectropolametric analysis}, we conducted an analysis of the $B_d$ assuming WD\,J1653$-$1001 has a simple dipole geometry. This yielded $B_d = 22^{+32}_{-7}$\,kG. Although this model provides a convenient first approximation, the true magnetic field geometry could be more complex. Further time-resolved spectropolarimetric monitoring will be needed to refine the constraints on the field geometry and increase the accuracy of $B_d$, as data were used in its calculation which we determine as non-detections. However, it seems likely that $B_d < 50$\,kG for WD\,J1653$-$1001. 

\subsection{Magnetic relation to antiphase variability}
\label{sec:magnetic relation to antiphase variability}
The $\langle B_z \rangle$ detections reported in this paper provide further constraints on the nature of WD\,J1653$-$1001. A significant detection of $\langle B_z \rangle = -9.16 \pm 2.39$\,kG in the 2024-07-15 FORS2 data occurred at $\phi = 0.51$, coinciding with low flux and strong Balmer line emission. The other significant detection, $\langle B_z \rangle = -4.78 \pm 1.23$\,kG, from the FORS2 data occurred on 2024-06-06 at $\phi = 0.86$, when the flux was increasing towards maximum, and the Balmer emission was nearing minimum strength. Similarly, the marginal detections found in the 2022-05-15 and 2022-06-01 FORS2 data occur at approximately the same phase as the 2024-06-06 FORS2 data: the former has $\langle B_z \rangle = -3.11 \pm 1.31$\,kG at $\phi = 0.86$; and the latter has $\langle B_z \rangle = -2.22 \pm 1.04$\,kG at $\phi = 0.93$. 

Taking into account the four FORS2 $\langle B_z \rangle$ detections, it seems that at large absolute $\langle B_z \rangle$ values, the star is at a phase of low photometric flux and strong emission. At small absolute $\langle B_z \rangle$ values, the star is at a phase of bright flux and weak emission. This is compelling evidence that if this star hosts a photospheric dark spot/region, it is not only temperature-inverted and has an optically thin chromospheric emission region, but is also magnetic.

\subsection{Spectral reclassification of \texorpdfstring{WD\,J1653$-$1001}{WDJ1653-1001}}
\label{sec:Spectral reclassification}
The detection of a weak magnetic field in WD\,J1653$-$1001 requires updating its spectral classification from DAe. The Balmer emission lines in this white dwarf are not visibly Zeeman-split, but do exhibit measurable polarisation.

Historically, the spectral type suffix `P' in white dwarf classifications has been used to indicate that magnetism was established through polarimetric detections, typically via continuum polarisation measurements. However, spectral lines affected by magnetic fields are inherently polarised, and whether they appear Zeeman-split (`H') or simply polarised (`P') depends on the spectral resolution and data quality. In the case of WD\,J1653$-$1001, the distinction between magnetism detected via polarimetry and that inferred from Zeeman line splitting is largely one of observational circumstance. At sufficiently high spectral resolution and S/N, the line profiles would likely show magnetic broadening consistent with a weak field. The essential physical property is the presence of a magnetic field in WD\,J1653$-$1001, not the particular diagnostic method used to reveal it.

Given these considerations, and to maintain consistency and clarity with the literature, we propose WD\,J1653$-$1001 is reclassified as a DAHe white dwarf. This designation conveys unambiguously that this object is a magnetic, hydrogen-dominated atmosphere white dwarf with emission lines, while avoiding potential confusion by the rarely used `P'. Therefore, WD\,J1653$-$1001 represents a low-field member of the DAHe class, whose magnetism is revealed primarily through polarimetry rather than spectroscopic Zeeman-splitting with the current data.

\subsection{Variability comparison with DA(H)e stars}
\label{sec:comparison}
Owing to a multitude of observational similarities and homogeneity in atmospheric parameters, we group the spectral classes DAHe and DAe together into one DA(H)e group. The 28 stars in this group have a likely common physical origin and mechanism causing variability. We now compare their characteristic photometric and spectroscopic variability and, where available, their magnetic field strength variability. 

For WD\,J1653$-$1001, the periodic antiphase relation is evident: phases of low photometric flux coincide with strong Balmer line emission and $\langle B_z \rangle$. The DAe  WD\,J0412$+$7549 \citep{Elms2023} exhibits the same photometric and spectroscopic antiphase variability. However, in the absence of detectable Zeeman-splitting and spectropolarimetry, the only magnetic field strength information known about WD\,J0412$+$7549 is an upper limit of $B < 0.05$\,MG. Time-resolved spectropolarimetry of WD\,J0412$+$7549 would therefore be crucial in determining whether it shares the same magnetic behavior as WD\,J1653$-$1001, so should be followed up in future. 

Among the DAHe stars in the literature, six have been studied in detail: GD\,356 \citep{Greenstein1985, Walters2021}, SDSS\,J1252$-$0234 \citep{Reding2020}, SDSS\,J1219$+$4715 \citep{Gansicke2020}, LP\,705$-$64 \citep{Reding2023}, WD\,J1430$-$5623 \citep{Reding2023}, and WD\,J1616$+$5410 \citep{Manser2023}. These stars all exhibit the same antiphase relation between broad-band photometry and strength of Balmer emission lines as WD\,J1653$-$1001 and WD\,J0412$+$7549. Furthermore, time-resolved magnetic field strength measurements reveal SDSS\,J1252$-$0234, LP\,705$-$64, WD\,J1430$-$5623 and WD\,J1616$+$5410 have the strongest fields at phases of low photometric flux and strong Balmer line emission - the same relation observed in WD\,J1653$-$1001. 

GD\,356, the first identified DAHe white dwarf \citep{Greenstein1985}, has a more complicated analysis. It has a relatively strong, near constant magnetic field strength of $B \approx 11.5$\,MG, and is the only DAHe to have spectropolarimetry \citep{Walters2021}. $\langle B_z \rangle$ measurements from WHT/ISIS show weak periodicity in potentially the opposite trend to other DAHe stars: $\langle B_z \rangle$ appears low when photometric flux is low and emission strength is high. However, the large uncertainties in the $\langle B_z \rangle$ measurements relative to the observed modulation amplitude in figure 8 of \citet{Walters2021} make this a very tentative correlation. The relatively high magnetic field strength of GD\,356 could have aligned the magnetic axis with the axis of rotation, distorting the time-resolved $\langle B_z \rangle$ measurements. Additional spectropolarimetric observations are required to determine whether GD\,356 truly deviates from the general DAHe pattern. 

SDSS\,J1219$+$4715 also appears to have a near constant magnetic field strength, with $B \approx 18.5$\,MG measured from Zeeman-splitting \citep{Gansicke2020}. However, it does not have spectropolarimetry to test for variability in $\langle B_z \rangle$. 

The remaining DAHe stars in the literature do not have adequate time-resolved spectroscopic nor spectropolarimetric data to determine whether they share the same antiphase variability of broad-band photometric flux, Balmer line emission and magnetic field strength observed in the above DA(H)e stars. Addressing this will require dedicated, phase-resolved spectroscopic and spectropolarimetric monitoring in future studies.

\subsection{Tentative detection of calcium}
\label{sec:tentative detection of Ca}
\begin{figure}
\centering
\includegraphics[width=\columnwidth]{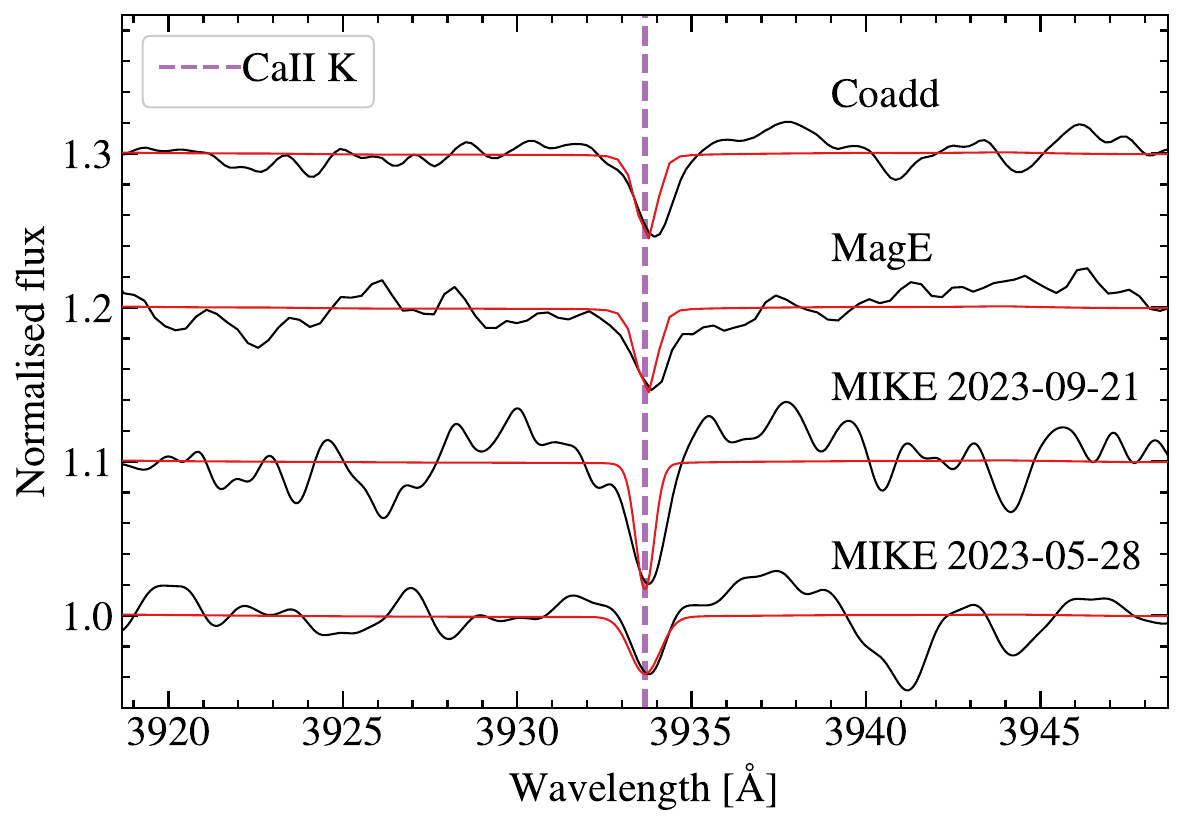}
    \caption{Observed spectra of WD\,J1653$-$1001 obtained with the MIKE and MagE instruments, together with a coadded spectrum, zoomed in on the \ion{Ca}{ii}~K (3934\,\AA) line. The UT dates of the two MIKE spectra are displayed. Model spectra are overlaid in red. All spectra are convolved with a Gaussian with a FWHM of 1\,\AA\ and offset vertically for clarity.}
\label{fig:Ca_Coadd_MIKE_MagE}
\end{figure}

DA(H)e white dwarfs, by the definition of their spectral type, do not display metal spectral lines. However, a faint \ion{Ca}{ii}~K (3934\,\AA) line was tentatively detected in the spectra of WD\,J1653$-$1001 from the MIKE and MagE instruments. We coadded the MIKE and MagE spectra and show the four spectra in Figure~\ref{fig:Ca_Coadd_MIKE_MagE}, which are normalised and zoomed on the \ion{Ca}{ii}~K region. Overlaid on the spectra are model spectra created from the latest updates of 3D DA local thermal equilibrium (LTE) atmosphere models from \citet{Tremblay2013} and \citet{Tremblay2015a}, with the addition of calcium. The \Teff, $\log g$, radius and distance of WD\,J1653$-$1001, listed in Table~\ref{tab:stellar parameters}, were used to create the model spectra. The calcium model abundances shown in Figure~\ref{fig:Ca_Coadd_MIKE_MagE} are: $-10.7$\,dex and $-10.5$\,dex for MIKE 2023-05-28 and 2023-09-21 spectra, respectively; $-10.1$\,dex for MagE; $-10.1$\,dex for the coadd. These are relatively low abundances of calcium which, if the detection is real, would likely have come from an earlier accretion event. Due to the relatively cool \Teff\ of WD\,J1653$-$1001, its atmosphere would be sufficiently dense for the accreted calcium to be observable on the order of Myr. We have found no evidence of active accretion on WD\,J1653$-$1001. 

If the main origin of this class is a single star magnetic phenomenon, there is likely no correlation with the presence of an evolved planetary system as seen in up to 50\% of all stellar remnants \citep{OuldRouis2024}. However, it may be more difficult to detect trace Ca abundances at the higher magnetic fields of DAHe white dwarfs.

The FORS2 2022 and Kast data also cover the \ion{Ca}{ii}~K region but the metal line was not detectable over the noise. The \ion{Ca}{ii}~H (3968\,\AA) line was not detected in any spectra, however this lies in the H$\epsilon$ Balmer line region so any low-abundance detection would be challenging. No other metal lines were identified in any spectra. 

We investigated the possibility that the tentative \ion{Ca}{ii}~K detection originates from the interstellar medium (ISM). WD\,J1653$-$1001 is relatively nearby at $32.63 \pm 0.04$\,pc and at Galactic latitude $b = 20.6$\,\degree. This corresponds to an approximate vertical height above the Galactic midplane of $z = d \mathrm{sin}b \approx 11.5$\,pc, where the ISM is relatively low density. This star has negligible reddening of $E(B-V) = 0.0056$ \citep{Vergely2022, Sahu2024}, which further indicates a very low ISM column density, rendering interstellar \ion{Ca}{ii}~K absorption unlikely.

We cannot confirm a \ion{Ca}{ii}~K detection at this time with our MagE and MIKE data alone. Additional spectra, ideally of high resolution, is required to confirm a detection and should be followed up at a later date.

\section{Conclusions}
\label{sec:Conclusions}
WD\,J1653$-$1001 has previously been classified as a DAe white dwarf, as it has H$\alpha$ and H$\beta$ line emission but does not exhibit Zeeman-splitting. The FORS2 spectropolarimetric observations presented and analysed in this work demonstrate that this star is weakly magnetic, with two significant line-of-sight magnetic field measurements of $\langle B_z \rangle = -9.2 \pm 2.4$\,kG and $\langle B_z \rangle = -4.8 \pm 1.2$\,kG, as well as two additional marginal detections. Magnetic fields of these strengths are insufficient to produce observable Zeeman-split spectral lines, hence the field previously went undetected with spectroscopy alone. We therefore propose that WD\,J1653$-$1001 is reclassified as a low-field DAHe white dwarf. 

Time-series analysis of the most recent publicly-available photometric data from ZTF and ATLAS, using two independent methods of MCMC and sinusoid fitting on the individual survey bands and combined datasets, provide robust evidence that WD\,J1653$-$1001 has a rotation period of $P \simeq 80.3$\,h. The best-fitting period from the simultaneous $g$-, $r$-, $c$- and $o$-band MCMC is $P = 80.3070 \pm 0.0007$\,h.

We also presented new and archival spectroscopic observations of WD\,J1653$-$1001 from Binospec, FORS2, MagE, INT, Kast and MIKE. Time-series analysis of the H$\alpha$ Balmer emission core \EW\ revealed a spectroscopic period of $P = 80.2922 \pm 0.0108$\,h. This is in agreement with the photometric period from ZTF and ATLAS. The concordance between these independent measurements validates $P \simeq 80.3$\,h as the rotation period of WD\,J1653$-$1001. 

We find the photometric flux and Balmer emission strength to vary in antiphase. The longitudinal magnetic field varies coherently with this cycle, with $\langle B_z \rangle$ reaching its strongest values at emission maxima and photometric flux minima. A likely mechanism causing this observed variability is a magnetically induced low optical flux spot/region in the stellar photosphere, producing an optically thin chromosphere responsible for Balmer emission \citep{Walters2021}.

Comparing WD\,J1653$-$1001 with the DAe star WD\,J0412$+$7549 and the growing sample of DAHe stars, reveals a consistent antiphase relationship: emission strength and magnetic field are strongest at phases of low photometric flux. This indicates that a unified physical mechanism is operating among DA(H)e stars. The similarities between DA(H)e white dwarfs, despite differing magnetic field strengths, highlights the importance of magnetic fields and possible surface inhomogeneities in characterising their variability. 

The results found in this work establish WD\,J1653$-$1001 as a benchmark system for understanding the complex interplay between magnetic fields, atmospheric structure and emission processes in white dwarfs. These findings bring the community one step closer to understanding the mechanisms causing these interesting and rare stars.

\section*{Acknowledgements}
This project has received funding from the European Research Council under the European Union’s Horizon 2020 research and innovation programme (Grant agreement numbers 101002408).

The observations from the FORS2 instrument were collected at the European Southern Observatory under ESO programme(s) 113.26ES.001.

This work has made use of data from the European Space Agency (ESA) mission
{\it Gaia} (\url{https://www.cosmos.esa.int/gaia}), processed by the {\it Gaia} Data Processing and Analysis Consortium (DPAC, \url{https://www.cosmos.esa.int/web/gaia/dpac/consortium}). Funding for the DPAC has been provided by national institutions, in particular the institutions participating in the {\it Gaia} Multilateral Agreement.

Based on observations obtained with the Samuel Oschin Telescope 48-inch and the 60-inch Telescope at the Palomar Observatory as part of the Zwicky Transient Facility project. ZTF is supported by the National Science Foundation under Grants No. AST-1440341 and AST-2034437 and a collaboration including current partners Caltech, IPAC, the Weizmann Institute for Science, the Oskar Klein Center at Stockholm University, the University of Maryland, Deutsches Elektronen-Synchrotron and Humboldt University, the TANGO Consortium of Taiwan, the University of Wisconsin at Milwaukee, Trinity College Dublin, Lawrence Livermore National Laboratories, IN2P3, University of Warwick, Ruhr University Bochum, Northwestern University and former partners the University of Washington, Los Alamos National Laboratories, and Lawrence Berkeley National Laboratories. Operations are conducted by COO, IPAC, and UW.

This work has made use of data from the Asteroid Terrestrial-impact Last Alert System (ATLAS) project. The Asteroid Terrestrial-impact Last Alert System (ATLAS) project is primarily funded to search for near earth asteroids through NASA grants NN12AR55G, 80NSSC18K0284, and 80NSSC18K1575; byproducts of the NEO search include images and catalogs from the survey area. This work was partially funded by Kepler/K2 grant J1944/80NSSC19K0112 and HST GO-15889, and STFC grants ST/T000198/1 and ST/S006109/1. The ATLAS science products have been made possible through the contributions of the University of Hawaii Institute for Astronomy, the Queen’s University Belfast, the Space Telescope Science Institute, the South African Astronomical Observatory, and The Millennium Institute of Astrophysics (MAS), Chile.

This work makes use of observations from the Las Cumbres Observatory global telescope network.

Research at Lick Observatory is partially supported by a generous gift from Google. A major upgrade of the Kast spectrograph on the Shane 3 m telescope at Lick Observatory was made possible through generous gifts from William and Marina Kast as well as the Heising-Simons Foundation.

The Isaac Newton Telescope is operated on the island of La Palma by the Isaac Newton Group of Telescopes in the Spanish Observatorio del Roque de los Muchachos of the Instituto de Astrofísica de Canarias.

This paper includes data gathered with the 6.5 meter Magellan Telescopes located at Las Campanas Observatory, Chile.

Observations reported here were obtained at the MMT Observatory, a joint facility of the Smithsonian Institution and the University of Arizona.

Based on observations collected at Centro Astronómico Hispano en Andalucía (CAHA) at Calar Alto, operated jointly by Junta de Andalucía and Consejo Superior de Investigaciones Científicas (IAA-CSIC).

\section*{Data Availability}
Spectroscopy from the VLT, INT and Shane telescope, in addition to photometry from ZTF, ATLAS and ASAS-SN, are available from their respective public archives. Photometry from the Calar Alto telescope is available upon reasonable request through astrohead@caha.es. Spectroscopy from Magellan and the MMT is available upon reasonable request. Spectropolarimetry from the VLT is available on the public archive (program IDs 109.235Q.002 and 113.26ES.001). 



\bibliographystyle{mnras}
\bibliography{all} 






\bsp	
\label{lastpage}
\end{document}